\documentclass[preprint,12pt]{elsarticle}

\usepackage[T1]{fontenc}
\usepackage{mathpazo}
\usepackage{microtype}
\usepackage{amsmath,amssymb,amsthm}
\usepackage{mathtools}
\usepackage{graphicx}
\usepackage{booktabs}
\usepackage[hidelinks]{hyperref}
\usepackage{orcidlink}

\usepackage[left=0.85in, right=0.85in, top=0.9in, bottom=0.9in, includefoot, headheight=13.6pt]{geometry}

\theoremstyle{plain}
\newtheorem{theorem}{Theorem}
\newtheorem{proposition}{Proposition}
\newtheorem{lemma}{Lemma}
\newtheorem{corollary}{Corollary}
\theoremstyle{definition}
\newtheorem{definition}{Definition}
\theoremstyle{remark}

\newcommand{\dd}{\widehat{e}}
\newcommand{\nn}{\widetilde{e}}
\newcommand{\Gc}{\Gamma_{\mathrm{c}}}
\DeclareMathOperator{\rank}{rank}
\DeclareMathOperator{\sign}{sign}
\DeclareMathOperator{\tr}{tr}

\journal{arXiv}

\begin{document}

\begin{frontmatter}

\title{Carrier relaxation confines the Bogdanov--Takens degeneracy of an optically injected semiconductor laser}

\author[1]{A. Martín-Ruiz\textsuperscript{\Large\orcidlink{0000-0001-5308-5448}}}
\author[2]{E. Chan-L\'opez\textsuperscript{\Large\orcidlink{0009-0003-7712-7907}\corref{cor1}}}
\author[2]{Carlos Antonio Mar\'in-Mendoza\textsuperscript{\Large\orcidlink{0009-0005-4821-0044}}}
\cortext[cor1]{Corresponding author: \texttt{eduardo.clopez13@gmail.com}}
\affiliation[1]{organization={Instituto de Ciencias Nucleares, Universidad Nacional Autónoma de México},
	city={Ciudad de México},
	postcode={04510},
	country={Mexico}}
\affiliation[2]{organization={División Académica de Ciencias Básicas, Universidad Juárez Autónoma de Tabasco},
            city={Cunduacán},
            postcode={86690},
            state={Tabasco},
            country={Mexico}}

\begin{abstract}
Optically injected semiconductor lasers are forced nonlinear oscillators whose locking boundaries are organized by codimension-two points. We ask which property of the carrier balance permits or forbids a Bogdanov--Takens (BT) point in the standard three-dimensional rate equations. Taking the carrier inversion $n$ as distinguished observable, the rate that enters the spectral conditions is the slope of the gain--loss balance, $\xi=\partial\dot n/\partial n=-2\Gamma/\kappa$, where $\Gamma$ is the damping of the relaxation oscillation and $\kappa=1+2Bn$ the normalized stimulated-emission factor. The two BT conditions become affine in $\xi$ and imply the exact sum rule $K^{2}=I[1-(1+\kappa)I]$ between injection amplitude $K$ and intracavity intensity $I$. The product of the two relaxation eigenvalues that survive at a saddle-node is bounded below by $1-16\Gamma^{2}$, so that no BT point exists for $\Gamma<1/4$, whatever $B$, the linewidth enhancement factor, the injection strength and the detuning; below this barrier every contact between Hopf and saddle-node curves is a fold--Hopf point. For $B=0$ the barrier is completed to an exact existence criterion. We prove that the unfolding of a BT point by injection strength and detuning is transversal if and only if two explicit curves cross transversally in the plane of locked states; for $B=0$ this is a simple root of an explicit function, and transversality fails exactly at the existence threshold. Two BT points at $\Gamma=1/2$, $\alpha=2$ are verified to be generic, and the contacts between Hopf and saddle-node curves are classified into fold--Hopf, BT and nilpotent triple-zero points.
\end{abstract}


\end{frontmatter}

%\linenumbers

%======================================================================
\section{Introduction}\label{sec:intro}
%======================================================================

A semiconductor laser subject to optical injection is a forced nonlinear oscillator whose low-dimensional rate equations have supported an unusually close dialogue between bifurcation theory and experiment \cite{Wieczorek2005,Wieczorek1999}. The intracavity field is pulled toward the injected frequency, the carrier reservoir feeds and depletes the field, and the linewidth enhancement factor couples the inversion to the optical frequency \cite{Henry1982}. The analytical tradition of the model goes back to explicit stability boundaries of the locked state \cite{Gavrielides1997} and to reduced equations that explain mechanisms such as period doubling \cite{Erneux1996}. Numerical continuation produced two-parameter bifurcation diagrams that agree quantitatively with measured stability maps over a wide range of injection strength and detuning \cite{Wieczorek2002}, and transitions near saddle-node--Hopf points were followed through the optical spectra of the device \cite{Wieczorek2003}.

The locking region is bounded by saddle-node and Hopf curves, and its codimension-two points separate qualitatively different regimes. For the parameter values usually taken as representative of semiconductor lasers, the Hopf and saddle-node curves touch at two saddle-node--Hopf (fold--Hopf) points \cite{Wieczorek2005,Wieczorek2003}. Beyond the locking boundary the dynamics is richer still: chaotic regions contain period-adding cascades that accumulate on smooth horizons \cite{Bonatto2007}; the route to phase locking passes through limit cycles of bounded phase whose character depends on how strongly the relaxation oscillation is damped \cite{Kelleher2012}; and excitable pulses at the boundary of the locking region of a quantum-dot laser show that this geometry has direct physical consequences \cite{Goulding2007}.

A Bogdanov--Takens (BT) point is the more degenerate possibility, in which the two spectral directions that become critical separately at a saddle-node and at a Hopf bifurcation collapse onto a double zero eigenvalue \cite{Bogdanov1975,Takens1974}. When it is generic, it is an organizing center from which a Hopf curve and a curve of homoclinic orbits emanate \cite{Kuznetsov2004}. The question addressed here is not how to locate such a point but whether the device permits one: \emph{which property of the physical balance of the laser permits or forbids the double zero eigenvalue?}

The answer is quantitative. The obstruction is visible in the free-running laser, whose linearization has the phase mode at zero and the relaxation-oscillation pair with product one in the natural units of the model. A saddle-node of a locked state keeps one zero eigenvalue; a BT point requires, in addition, that the product of the remaining pair vanish. We show that the carrier balance bounds this product from below by $1-16\Gamma^{2}$, where $\Gamma$ is the damping of the relaxation oscillation measured in units of its frequency, and that the bound is the shadow of an exact sum rule satisfied at every BT point. The device physics acts as a confinement of the degeneracy: there is a region of parameter space where the BT point is permitted and a region where the balance equations forbid it.

The analysis rests on one observation. With the carrier inversion $n$ as observable, the characteristic invariants of the Jacobian are affine in a single scalar $\xi$, and at a locked state $\xi$ is the slope of the gain--loss balance with respect to the inversion. The scalar that enters the spectral reduction is a physical rate fixed before any eigenvalue is computed. This turns the BT conditions from two polynomial equations into two statements about the balance law, and those statements produce inequalities.

Existence of the degenerate equilibrium is not the whole story. A BT point organizes the standard local bifurcation diagram only if the quadratic normal-form coefficients do not vanish and the two parameters unfold the singularity transversally \cite{Kuznetsov2004}. We prove that, for injection strength and detuning as unfolding parameters, transversality is equivalent to the transversal crossing of the curves $e_{2}=0$ and $e_{3}=0$ in the plane of locked states. For a gain law without the inversion dependence measured by $B$, it reduces to a simple root of an explicit function, and it fails exactly where pairs of BT points are born. We also classify the contacts between Hopf and saddle-node curves, which can be fold--Hopf points, BT points or nilpotent triple-zero points.

The paper is organized as follows. Section~\ref{sec:model} recalls the model and the parametrization of locked states. Section~\ref{sec:xi} introduces the transport rate $\xi$ and the closed forms of the characteristic invariants. Section~\ref{sec:BT} states the BT conditions and the four conditions of genericity. Sections~\ref{sec:sum}--\ref{sec:sharp} derive the sum rule, the relaxation barrier and the exact existence criterion for $B=0$. Section~\ref{sec:transv} proves the transversality criterion. Section~\ref{sec:contacts} classifies the contacts between Hopf and saddle-node curves. Section~\ref{sec:geometry} describes the bifurcation geometry and the periodic orbits near one BT point, and Sec.~\ref{sec:example} constructs and verifies two generic BT points. Section~\ref{sec:physics} separates exact results, physical interpretation, testable predictions and open questions, and Sec.~\ref{sec:conclusions} concludes. The algebra is collected in \ref{app:A}. Throughout, analytical statements come first; numerical computations are used to visualize them, to test them independently, or, where stated, to evaluate quantities at specific parameter values.

%======================================================================
\section{The injected laser as a three-dimensional balance}\label{sec:model}
%======================================================================

We use the rate equations of a single-mode semiconductor laser with optical injection in the form of Ref.~\cite[Eq.~(65)]{Wieczorek2005},
\begin{subequations}\label{eq:model}
\begin{align}
\dot E&=K+\Bigl[\tfrac12(1+i\alpha)\,n-i\omega\Bigr]E,\label{eq:modelE}\\
\dot n&=-2\Gamma n-(1+2Bn)\bigl(|E|^{2}-1\bigr).\label{eq:modeln}
\end{align}
\end{subequations}
Here $E=x+iy$ is the slowly varying intracavity field in the frame of the injected light, normalized so that the free-running intensity $I=|E|^{2}$ equals one; $n$ is the carrier inversion measured from its threshold value; $K\ge0$ is the injected field strength and $\omega$ the detuning between injected and free-running frequencies; $\alpha$ is the linewidth enhancement factor \cite{Henry1982}. Time is measured in units of the inverse relaxation-oscillation frequency $\omega_{R}^{-1}$. In terms of the carrier and photon decay rates $\gamma_{N}$ and $\gamma_{E}$, the two remaining parameters are \cite{Wieczorek2005}
\begin{equation}\label{eq:scaling}
B=\frac{\omega_{R}}{2\gamma_{E}},\qquad
\Gamma=\frac12\Bigl(\frac{\gamma_{N}}{\omega_{R}}+\frac{\omega_{R}}{\gamma_{E}}\Bigr)=\frac{\gamma_{N}}{2\omega_{R}}+B .
\end{equation}
We write
\begin{equation}\label{eq:kappa}
\kappa:=1+2Bn .
\end{equation}
The factor $\kappa$ multiplies the stimulated term $|E|^{2}-1$ in \eqref{eq:modeln}: it is the stimulated-emission rate per unit intensity at inversion $n$, normalized to its value at threshold, and $B$ measures its dependence on the inversion. We stress that $B$ is not a gain-compression (gain-saturation) coefficient; in the scaling \eqref{eq:scaling} it is the ratio of the relaxation-oscillation frequency to twice the photon decay rate. Physical states have $\kappa>0$. By \eqref{eq:scaling}, $\Gamma-B=\gamma_{N}/(2\omega_{R})>0$, so that
\begin{equation}\label{eq:GB}
\Gamma>B\ge\tfrac12 B
\end{equation}
for every device described by \eqref{eq:model}.

The three real variables $(x,y,n)$ are three distinct physical degrees of freedom: the two quadratures of the field and the reservoir that feeds it. Equation~\eqref{eq:modeln} is a balance law: pumping above transparency against spontaneous relaxation at rate $2\Gamma$ and stimulated depletion by the field. The phase space is three-dimensional, and this is what allows a double zero eigenvalue to coexist with a third, transverse eigenvalue.

\emph{Free-running spectrum.} For $K=0$ the state $E=1$, $n=0$ has the characteristic polynomial $\lambda(\lambda^{2}+2\Gamma\lambda+1)$. The zero eigenvalue is the phase (Goldstone) mode, and the pair $\lambda=-\Gamma\pm i(1-\Gamma^{2})^{1/2}$ is the relaxation oscillation, with product one and damping $\Gamma$. In physical units its damping rate is $\Gamma\omega_{R}=\frac12(\gamma_{N}+\omega_{R}^{2}/\gamma_{E})$. Everything that follows measures how far injection can move this pair.

\emph{Locked states.} An equilibrium with $E\neq0$ is a locked state. We parametrize locked states by the inversion $n$ and by the effective detuning
\begin{equation}\label{eq:s}
s:=\omega-\tfrac12\alpha n ,
\end{equation}
the detuning seen by the field once the carrier-induced frequency shift is subtracted. The field equation gives $E=-K/(\frac12n-is)$, and $\dot n=0$ gives the intensity as a function of the inversion alone:
\begin{equation}\label{eq:balance}
I=1-\frac{2\Gamma n}{\kappa},\qquad
K^{2}=I\,d,\qquad d:=\frac{n^{2}}{4}+s^{2}.
\end{equation}
At fixed $(K,\omega)$ the locked states are the roots in $n$ of the locking function
\begin{equation}\label{eq:locking}
F(n;K,\omega)=I(n)\Bigl[\frac{n^{2}}{4}+\Bigl(\omega-\frac{\alpha n}{2}\Bigr)^{2}\Bigr]-K^{2},
\end{equation}
and $\kappa F$ is a cubic polynomial in $n$, so there are at most three locked states. Conversely, every pair $(n,s)$ with $I(n)>0$, $\kappa(n)>0$ and $(n,s)\neq(0,0)$ defines a unique locked state together with the injection parameters
\begin{equation}\label{eq:chart}
(\omega,K)=\bigl(s+\tfrac12\alpha n,\ \sqrt{I(n)\,d}\,\bigr)
\end{equation}
that sustain it. The map $(n,s)\mapsto(x,y,n,K,\omega)$ is smooth with smooth inverse $(x,y,n,K,\omega)\mapsto(n,\omega-\alpha n/2)$; it is a global chart of the surface of locked states with $K>0$ in the five-dimensional space of states and injection parameters. This inversion of the usual viewpoint, from parameters to states and back, is used throughout.

%======================================================================
\section{The transport rate of the carrier balance}\label{sec:xi}
%======================================================================

\subsection{Definition and physical meaning}

Let $J$ denote the Jacobian of \eqref{eq:model} in the real coordinates $(x,y,n)$ at a locked state, and let $e_{1},e_{2},e_{3}$ be its elementary characteristic invariants, $\det(\lambda-J)=\lambda^{3}-e_{1}\lambda^{2}+e_{2}\lambda-e_{3}$, so that $e_{1}=\tr J$, $e_{3}=\det J$ and $e_{2}$ is the sum of the principal $2\times2$ minors.

The model distinguishes one observable, the inversion $h=n$, and one direction transverse to its level sets, $u=\partial_{n}$, the change of inversion at fixed field. We call
\begin{equation}\label{eq:xidef}
\xi:=D_{u}(\dot h)=\frac{\partial\dot n}{\partial n}=-2\Gamma-2B\,(I-1)
\end{equation}
the transport rate of the carrier balance: the rate at which the balance responds when carriers are added at fixed field. At an equilibrium it coincides with the entry $J_{33}$ of the Jacobian; the point of the definition is that it is a property of the balance law, not of the matrix.

\begin{lemma}[Transport rate on the balance surface]\label{lem:xi}
At every locked state,
\begin{equation}\label{eq:xi}
\xi=-\frac{2\Gamma}{\kappa},\qquad I=1+\xi n,\qquad
\frac{dI}{dn}\Big|_{\dot n=0}=\frac{\xi}{\kappa},\qquad
\frac{d^{2}I}{dn^{2}}\Big|_{\dot n=0}=\frac{8\Gamma B}{\kappa^{3}} .
\end{equation}
\end{lemma}

\begin{proof}
On $\dot n=0$, $I-1=-2\Gamma n/\kappa$; inserting this into \eqref{eq:xidef} gives $\xi=-2\Gamma(\kappa-2Bn)/\kappa=-2\Gamma/\kappa$. The remaining identities follow by differentiating $I(n)=1-2\Gamma n/\kappa$.
\end{proof}

The quantity $\xi$ is fixed by the device before any eigenvalue is computed: for $B=0$ it is the constant $-2\Gamma$, and in general it is $\kappa$ times the slope of the balance curve in the $(n,I)$ plane. That curve is a straight line through the free-running point $(0,1)$ when $B=0$, and it bends with curvature proportional to $B$ otherwise (Fig.~\ref{fig:balance}). Physically, $-\xi$ is the rate at which an excess of carriers created at fixed optical field relaxes back to the balance: spontaneous relaxation contributes $2\Gamma$ and the inversion dependence of stimulated emission contributes $2B(I-1)$, which on the balance surface combine into $2\Gamma/\kappa$.

\subsection{Invariants in the gain--loss frame}

Removing $\xi$ from the Jacobian splits it into two pieces. The reduced matrix $\widehat J$ is the field block, the restriction of $J$ to the field quadratures at fixed inversion; the corrected matrix $\widetilde J$ is $J$ with the entry $J_{33}=\xi$ replaced by zero. We write $\dd_{k}=e_{k}(\widehat J)$ and $\nn_{k}=e_{k}(\widetilde J)$.

\begin{proposition}[Closed forms]\label{prop:closed}
At every locked state,
\begin{equation}\label{eq:closed}
\dd_{1}=n,\qquad \dd_{2}=d=\frac{K^{2}}{I},\qquad
\nn_{2}=d+\kappa I,\qquad
\nn_{3}=\kappa I\Bigl(\frac n2-\alpha s\Bigr),
\end{equation}
and the invariants of $J$ are affine in $\xi$:
\begin{equation}\label{eq:border}
e_{1}=n+\xi,\qquad e_{2}=\xi\dd_{1}+\nn_{2},\qquad e_{3}=\xi\dd_{2}+\nn_{3}.
\end{equation}
Explicitly, using $\xi n=I-1$,
\begin{equation}\label{eq:e23}
e_{2}=d+(1+\kappa)I-1,\qquad
e_{3}=\xi d+\kappa I\Bigl(\frac n2-\alpha s\Bigr).
\end{equation}
\end{proposition}

\begin{proof}
In real coordinates
\begin{equation}\label{eq:J}
J=\begin{pmatrix}
n/2 & s & \tfrac12(x-\alpha y)\\
-s & n/2 & \tfrac12(\alpha x+y)\\
-2\kappa x & -2\kappa y & \xi
\end{pmatrix},
\end{equation}
whose field block has trace $n$ and determinant $d$, which equals $K^{2}/I$ by \eqref{eq:balance}. Every principal minor of $J$ that contains the $(3,3)$ entry is affine in it, with coefficient the complementary minor of the field block; this gives \eqref{eq:border}. The remaining terms are computed in \ref{app:inv}.
\end{proof}

\begin{lemma}[Saddle-node as a fold of the locking function]\label{lem:fold}
At every locked state, $\det J=\kappa\,\partial_{n}F(n;K,\omega)$, with $F$ given by \eqref{eq:locking}.
\end{lemma}

\begin{proof}
By \eqref{eq:locking} and Lemma~\ref{lem:xi}, $\partial_{n}F=(\xi/\kappa)d+I(n/2-\alpha s)$; multiplying by $\kappa$ gives \eqref{eq:e23}.
\end{proof}

A saddle-node is therefore exactly a double root of the locking equation, and the slope $\xi$ of the balance enters it through the first term of $e_{3}$.

%======================================================================
\section{Bogdanov--Takens points and their genericity}\label{sec:BT}
%======================================================================

\begin{definition}\label{def:BT}
A locked state is a BT point if $e_{3}=e_{2}=0$, $e_{1}\neq0$ and $\rank J=2$.
\end{definition}

The rank condition is automatic: the field block has determinant $d=K^{2}/I>0$, so $\rank J\ge2$ at every locked state with $K>0$. At a BT point the spectrum is $\{0,0,\lambda_{3}\}$ with $\lambda_{3}=e_{1}=n+\xi\neq0$, and the double zero eigenvalue has a single Jordan block. The BT conditions do not fix the sign of the transverse eigenvalue. If $\lambda_{3}<0$, that is $n<-\xi$, the transverse direction is attracting; if $\lambda_{3}>0$, that is $n>-\xi$, it is repelling. Both cases occur (Sec.~\ref{sec:contacts}). In either case it is $\lambda_{3}\neq0$ that permits the reduction to a two-dimensional parameter-dependent center manifold, on which the BT analysis is carried out; the sign of $\lambda_{3}$ only decides whether the planar dynamics on that manifold attracts nearby orbits of the full system ($\lambda_{3}<0$) or is of saddle type in the three-dimensional phase space ($\lambda_{3}>0$).

By \eqref{eq:border}, the two BT conditions read
\begin{equation}\label{eq:BTaffine}
\xi\,\dd_{2}+\nn_{3}=0,\qquad \xi\,\dd_{1}+\nn_{2}=0 ,
\end{equation}
two equations affine in the single scalar $\xi$. The consequence $\xi=-\nn_{3}/\dd_{2}=-\nn_{2}/\dd_{1}$ is an algebraic rewriting of \eqref{eq:BTaffine} and carries no information by itself. The content appears when $\xi$ is replaced by the value that the balance law imposes, Lemma~\ref{lem:xi}: the left-hand sides then become functions of the state, and the equations become constraints on where a BT point can sit.

Definition~\ref{def:BT} concerns the equilibrium only. Following Ref.~\cite[Secs.~8.4 and 8.7]{Kuznetsov2004}, a BT point of a two-parameter family is \emph{generic} if, in addition to (G0) the spectral condition of Definition~\ref{def:BT} with $\lambda_{3}\neq0$, the following hold. Let $q_{0},q_{1}$ and $p_{0},p_{1}$ be the right and left Jordan chains of $J$, normalized as in \ref{app:normal}, and let $a$ and $b$ be the coefficients of the normal form $\dot w_{0}=w_{1}$, $\dot w_{1}=aw_{0}^{2}+bw_{0}w_{1}$ on the center manifold. Then:
\begin{itemize}
\item[(G1)] $a\neq0$;
\item[(G2)] $b\neq0$;
\item[(G3)] (transversality) on the parameter-dependent center manifold, with local coordinate $u\in\mathbb R^{2}$ and reduced vector field $f_{c}(u;K,\omega)$, the map $(u,K,\omega)\mapsto\bigl(f_{c},\ \tr D_{u}f_{c},\ \det D_{u}f_{c}\bigr)$ is regular at the BT point.
\end{itemize}
Conditions (G1)--(G2) concern the germ of the vector field at the point; (G3) concerns the way the two parameters move it. Under (G0)--(G3) the local bifurcation diagram in the $(\omega,K)$ plane is the standard one: a saddle-node curve, a Hopf curve and a curve of homoclinic orbits emanate from the point, and the sign of $ab$ decides whether the Hopf curve and the periodic orbits near the point are supercritical ($ab<0$) or subcritical ($ab>0$) \cite{Kuznetsov2004}. These statements refer to the dynamics on the center manifold; when $\lambda_{3}>0$ the corresponding orbits are of saddle type in the full system. Sections~\ref{sec:sum}--\ref{sec:sharp} concern (G0); Sec.~\ref{sec:transv} reduces (G3) to an explicit condition; (G1)--(G2) are evaluated at specific points in Sec.~\ref{sec:example}.

%======================================================================
\section{The sum rule}\label{sec:sum}
%======================================================================

\begin{theorem}[Sum rule]\label{thm:sum}
At every BT point of \eqref{eq:model},
\begin{equation}\label{eq:sumrule}
K^{2}=I\bigl[1-(1+\kappa)I\bigr].
\end{equation}
Consequently
\begin{equation}\label{eq:bounds}
I<\frac{1}{1+\kappa},\qquad K\le\frac{1}{2\sqrt{1+\kappa}},\qquad 0<n<2 .
\end{equation}
\end{theorem}

\begin{proof}
The condition $e_{2}=0$ in the form \eqref{eq:e23} is $K^{2}/I=1-(1+\kappa)I$, which is \eqref{eq:sumrule}. Since $K^{2}/I=d\ge0$, it forces $(1+\kappa)I\le1$, with equality only for $n=s=0$, hence $K=0$, which is not a locked state under injection. The function $I\mapsto I[1-(1+\kappa)I]$ has maximum $1/[4(1+\kappa)]$, which bounds $K$. Finally $n^{2}/4\le d=1-(1+\kappa)I<1$ gives $|n|<2$, and $n\le0$ is excluded by Theorem~\ref{thm:barrier} below, whose proof shows $e_{2}>0$ for $n\le0$.
\end{proof}

\begin{figure}[t]
\centering
\includegraphics[width=\textwidth]{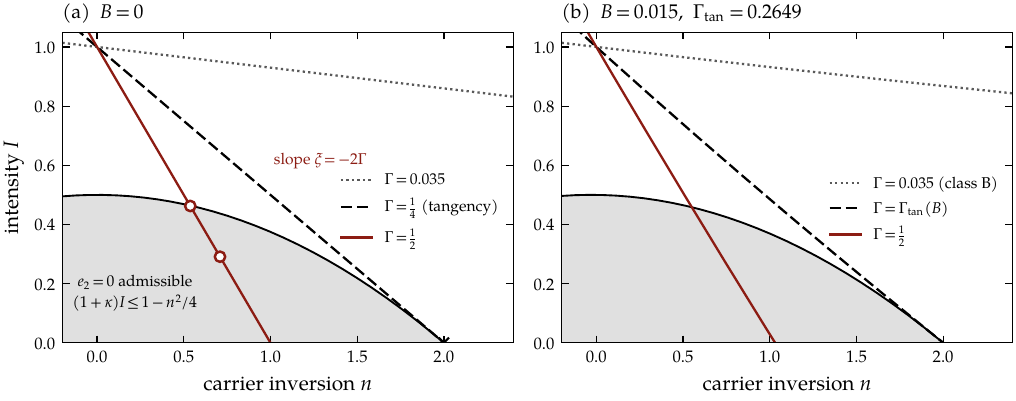}
\caption{Balance geometry. Shaded: the region $\mathcal R_{B}$ of Eq.~\eqref{eq:region}, where $e_{2}=0$ is compatible with a locked state; every BT point lies in it. Lines: the gain--loss balance $I=1-2\Gamma n/\kappa$, whose slope at the free-running point is $\xi$. (a)~$B=0$: $\mathcal R_{0}$ is bounded by the parabola $I=\frac12-n^{2}/8$, and the balance line meets it if and only if $\Gamma\ge\frac14$, with tangency at $(n,I)=(2,0)$. Open circles: the two BT points of Sec.~\ref{sec:example} ($\Gamma=\frac12$, $\alpha=2$); for fixed $\Gamma$ and varying $\alpha$ the BT points move along the chord inside $\mathcal R_{0}$. (b)~$B=0.015$; the class-B line $\Gamma=0.035$ stays far from $\mathcal R_{B}$; dashed: the $\alpha$-independent threshold $\Gamma_{\tan}(B)$ of Proposition~\ref{prop:tan}.}
\label{fig:balance}
\end{figure}

The sum rule is exact and contains no parameter beyond $B$, through $\kappa$. It uses only $e_{2}=0$ and the balance law; the condition $e_{3}=0$ has not been used, which is why neither $\alpha$ nor $\omega$ appears. Its physical content is that a BT point cannot occur at high intensity: the injected laser must be driven into a state in which the intracavity intensity is depleted below $1/(1+\kappa)$ of its free-running value, about one half when $B$ is small, and the injection cannot exceed $2^{-3/2}\approx0.354$ in the same limit. In the $(n,I)$ plane, BT points lie in the region
\begin{equation}\label{eq:region}
\mathcal R_{B}=\Bigl\{(n,I):\ I>0,\ (1+\kappa)\,I\le1-\frac{n^{2}}{4}\Bigr\},
\end{equation}
which is the projection of $\{e_{2}=0\}$, and on the balance curve \eqref{eq:balance}. The barrier of the next section is the statement that the balance curve can fail to reach $\mathcal R_{B}$ (Fig.~\ref{fig:balance}).

%======================================================================
\section{Relaxation as a spectral barrier}\label{sec:barrier}
%======================================================================

On the balance surface, \eqref{eq:e23} and \eqref{eq:balance} give
\begin{equation}\label{eq:e2q}
e_{2}=q(n)+s^{2},\qquad
q(n):=\frac{n^{2}}{4}+\kappa-\frac{2\Gamma n(1+\kappa)}{\kappa}.
\end{equation}
Since $s$ enters only through $s^{2}$, the parameters $\alpha$, $K$ and $\omega$, which act on the state only through $s$, drop out of any lower bound on $e_{2}$.

\begin{theorem}[Relaxation barrier]\label{thm:barrier}
Let $B\ge0$ and $\Gamma\ge B/2$. At every locked state of \eqref{eq:model},
\begin{equation}\label{eq:gap}
e_{2}\ \ge\ 1-16\,\Gamma^{2}.
\end{equation}
Moreover, for every $B\ge0$ (without restriction on $\Gamma$), $\alpha$, $K$ and $\omega$, a BT point requires
\begin{equation}\label{eq:quarter}
\Gamma\ \ge\ \tfrac14 .
\end{equation}
\end{theorem}

\begin{proof}
For $n\ge0$ one has $\kappa\ge1$, and a direct computation gives
\begin{equation}\label{eq:qgap}
q(n)-\Bigl(\frac{n^{2}}{4}-4\Gamma n+1\Bigr)
=\frac{2Bn\,(2Bn+2\Gamma n+1)}{2Bn+1}\ \ge\ 0 ,
\end{equation}
so $q(n)\ge n^{2}/4-4\Gamma n+1\ge1-16\Gamma^{2}$, the last quadratic being minimal at $n=8\Gamma$. For $n<0$ with $\kappa>0$, the term $-2\Gamma n(1+\kappa)/\kappa$ is positive, so $q(n)>n^{2}/4+\kappa=n^{2}/4+1+2Bn\ge1-4B^{2}$. Hence $e_{2}\ge\min\{1-16\Gamma^{2},1-4B^{2}\}$ by \eqref{eq:e2q}, which is \eqref{eq:gap} when $\Gamma\ge B/2$. For the barrier, let $\Gamma<\frac14$ and $B\ge0$ be arbitrary: for $n\ge0$, $e_{2}\ge1-16\Gamma^{2}>0$, and for $n<0$, $e_{2}\ge q(n)>n^{2}/4+\kappa>0$. Thus $e_{2}=0$ is impossible.
\end{proof}

By \eqref{eq:GB} the hypothesis $\Gamma\ge B/2$ holds for every device described by \eqref{eq:model}, and the barrier \eqref{eq:quarter} does not need it.

Theorem~\ref{thm:barrier} is the central physical result. Its reading is spectral. On the saddle-node set $e_{3}=0$, one eigenvalue is zero and $e_{2}=\lambda_{2}\lambda_{3}$ is the product of the remaining two, the continuation of the relaxation-oscillation pair whose product equals one in the free-running laser. Inequality \eqref{eq:gap} says that injection can lower this product by at most $16\Gamma^{2}$. A BT point is the event in which one member of the pair is driven to the phase mode at zero; when $\Gamma<\frac14$ the carrier balance does not allow it. Carrier relaxation acts as a spectral barrier for the degeneracy.

Two remarks delimit the statement. First, \eqref{eq:quarter} is a necessary condition and is not sharp at fixed $\alpha$: Sec.~\ref{sec:sharp} shows that for $B=0$ the actual threshold lies strictly above $\frac14$ for every finite $\alpha$. Second, \eqref{eq:gap} is a statement about the model \eqref{eq:model}; its relation to devices is discussed in Sec.~\ref{sec:physics}.

\begin{corollary}[Class-B regime]\label{cor:classB}
For the values $\Gamma=0.035$, $B=0.015$ used in Ref.~\cite{Wieczorek2005} as representative of a semiconductor laser pumped at $1.5$ times threshold, $e_{2}\ge0.9804$ at every locked state; no BT point exists for any $\alpha$, $K$, $\omega$.
\end{corollary}

A direct evaluation along the saddle-node curves gives minima of $e_{2}$ equal to $0.9939$, $0.9904$, $0.9892$, $0.9885$ and $0.9882$ for $\alpha=1,2,3,5,8$; for $\alpha=0$ the values stay above $1$ and approach $1$ in the weak-injection limit $n\to0$ (numerical evaluation). These numbers are consistent with \eqref{eq:gap} and show that it is nearly attained.

The quarter in \eqref{eq:quarter} is the $B=0$ value of an exact $\alpha$-independent threshold.

\begin{proposition}[Tangency threshold]\label{prop:tan}
For $B\ge0$ let
\begin{equation}\label{eq:gtan}
\Gamma_{\tan}(B):=\inf_{0<n<2}\ \frac{\kappa\,(n^{2}/4+\kappa)}{2n(1+\kappa)} .
\end{equation}
A BT point requires $\Gamma\ge\Gamma_{\tan}(B)$, and $\frac14=\Gamma_{\tan}(0)\le\Gamma_{\tan}(B)\le\frac14(1+4B)$.
\end{proposition}

\begin{proof}
At a BT point $q(n)=-s^{2}\le0$ with $0<n<2$ by Theorem~\ref{thm:sum}; for $n>0$, $q(n)\le0$ is equivalent to $\Gamma\ge\kappa(n^{2}/4+\kappa)/[2n(1+\kappa)]$. The bounds follow from \eqref{eq:qgap} and from the value $(1+4B)/4$ at $n=2$.
\end{proof}

Geometrically, $\Gamma_{\tan}(B)$ is the damping at which the balance curve first touches the region $\mathcal R_{B}$ (Fig.~\ref{fig:balance}). For $B=0.015$ one finds $\Gamma_{\tan}=0.2649$, and for $B=0.1$, $\Gamma_{\tan}=0.3472$.

%======================================================================
\section{The exact threshold for \texorpdfstring{$B=0$}{B=0}}\label{sec:sharp}
%======================================================================

For $B=0$ the BT conditions reduce to one equation in one unknown. Put
\begin{equation}\label{eq:sigmag}
\sigma(n)=4\Gamma n-1-\frac{n^{2}}{4},\qquad
g(n)=\frac n2-\frac{2\Gamma(4\Gamma n-1)}{1-2\Gamma n},
\end{equation}
and, for $\Gamma>\frac14$, let $n_{-}=8\Gamma-2(16\Gamma^{2}-1)^{1/2}$ be the smaller root of $\sigma$. With $I=1-2\Gamma n$, the condition $e_{2}=0$ is $s^{2}=\sigma(n)$, and, given it, $e_{3}=0$ is $\alpha s=g(n)$; more precisely, for $B=0$ the identity
\begin{equation}\label{eq:e3B0}
e_{3}=I\,\bigl(g(n)-\alpha s\bigr)-2\Gamma\,e_{2}
\end{equation}
holds at every locked state (\ref{app:sharp}). Hence the BT points with $\alpha>0$ are in one-to-one correspondence with the roots of
\begin{equation}\label{eq:Phi}
\Phi_{\alpha}(n)=g(n)^{2}-\alpha^{2}\sigma(n)
\end{equation}
in the window $W_{\Gamma}=[n_{-},1/(2\Gamma))$, with $s=g/\alpha$. For $\alpha=0$ the condition $e_{3}=0$ reduces to $g(n)=0$, which is again $\Phi_{0}(n)=g(n)^{2}=0$, and $s=\pm\sigma(n)^{1/2}$. The window is nonempty exactly when $\Gamma>\frac14$, which recovers \eqref{eq:quarter}. Since $(\alpha,s,\omega)\mapsto(-\alpha,-s,-\omega)$ maps BT points to BT points, we take $\alpha\ge0$.

\begin{theorem}[Existence criterion, $B=0$]\label{thm:sharp}
Let $B=0$, $\Gamma>\frac14$ and
\begin{equation}\label{eq:alpha0}
\alpha_{0}(\Gamma):=\inf_{n_{-}<n<1/(2\Gamma)}\ \frac{|g(n)|}{\sqrt{\sigma(n)}} .
\end{equation}
Equations $e_{2}=e_{3}=0$ have a solution with $K>0$, $I>0$ if and only if $\alpha\ge\alpha_{0}(\Gamma)$. Moreover:
\begin{itemize}
\item[(i)] $\alpha_{0}(\Gamma)=0$ if and only if $\Gamma\ge\Gc:=1/(2\sqrt3)$: for these damping rates BT conditions are met for every $\alpha$, and for $\alpha=0$ they are met exactly for these damping rates;
\item[(ii)] $\alpha_{0}(\Gamma)<\infty$ for every $\Gamma>\frac14$, and $\alpha_{0}(\Gamma)\to\infty$ as $\Gamma\to\frac14^{+}$.
\end{itemize}
The solutions are BT points except where $e_{1}=n-2\Gamma$ vanishes, which happens only on the triple-zero curve
\begin{equation}\label{eq:TZ}
\alpha_{\mathrm{TZ}}(\Gamma)=\frac{\Gamma\,|3-20\Gamma^{2}|}{(1-4\Gamma^{2})\sqrt{7\Gamma^{2}-1}},
\qquad \frac{1}{\sqrt7}<\Gamma<\frac12 .
\end{equation}
\end{theorem}

\begin{proof}
On $W_{\Gamma}$ one has $I>0$, and $\sigma\ge0$ because $\sigma(1/(2\Gamma))=1-1/(16\Gamma^{2})>0$, so that $1/(2\Gamma)$ lies between the roots of $\sigma$; moreover $\sigma>0$ on the interior $W_{\Gamma}^{\circ}=(n_{-},1/(2\Gamma))$ and $\sigma(n_{-})=0$. Since $n\ge n_{-}>0$, every solution has $K^{2}=I(\sigma+n^{2}/4)>0$. By \eqref{eq:e3B0} and $e_{2}=s^{2}-\sigma$, the conditions $e_{2}=e_{3}=0$ with $I>0$ are equivalent to $n\in W_{\Gamma}$, $s^{2}=\sigma(n)$ and $\alpha s=g(n)$. For $\alpha>0$ this is $\Phi_{\alpha}(n)=0$ with $s=g/\alpha$; for $\alpha=0$ it is $g(n)=0$ with $s=\pm\sigma(n)^{1/2}$, that is, again $\Phi_{0}(n)=0$. In every case a solution exists if and only if $\Phi_{\alpha}$ has a zero in $W_{\Gamma}$.

Let $Q=|g|/\sqrt\sigma$ on $W_{\Gamma}^{\circ}$, where $\Phi_{\alpha}=\sigma\,(Q^{2}-\alpha^{2})$. As $n\to1/(2\Gamma)^{-}$, $g\to-\infty$ while $\sigma$ stays bounded and positive, so $Q\to+\infty$ and $\Phi_{\alpha}\to+\infty$.

If $\Phi_{\alpha}(n_{*})=0$ with $n_{*}\in W_{\Gamma}^{\circ}$, then $Q(n_{*})=\alpha$ and $\alpha\ge\alpha_{0}(\Gamma)$. If $n_{*}=n_{-}$, then $\sigma(n_{-})=0$ forces $g(n_{-})=0$; since $\sigma'(n_{-})=(16\Gamma^{2}-1)^{1/2}>0$ and $g$ is smooth at $n_{-}<1/(2\Gamma)$, $Q(n)=O\bigl((n-n_{-})^{1/2}\bigr)$ as $n\to n_{-}^{+}$, so $\alpha_{0}(\Gamma)=0\le\alpha$.

Conversely, if $\alpha>\alpha_{0}(\Gamma)$ there is $n_{1}\in W_{\Gamma}^{\circ}$ with $Q(n_{1})<\alpha$, so $\Phi_{\alpha}(n_{1})<0$, and $\Phi_{\alpha}$ vanishes in $(n_{1},1/(2\Gamma))$ by continuity. If $\alpha=\alpha_{0}(\Gamma)$ and the infimum is attained at $n_{0}\in W_{\Gamma}^{\circ}$, then $\Phi_{\alpha}(n_{0})=0$. If $\alpha=\alpha_{0}(\Gamma)$ and the infimum is not attained, a minimizing sequence accumulates at $n_{-}$, because $Q\to+\infty$ at the right end; if $g(n_{-})\neq0$, then $Q\to+\infty$ at $n_{-}$ as well, a contradiction; hence $g(n_{-})=0$ and $\Phi_{\alpha}(n_{-})=g(n_{-})^{2}-\alpha^{2}\sigma(n_{-})=0$. This proves the criterion for every $\alpha\ge0$, including $\alpha=0$ and the equality case $\alpha=\alpha_{0}(\Gamma)$. Items (i) and (ii) and the triple-zero curve are proved in \ref{app:sharp}.
\end{proof}

\begin{figure}[t]
\centering
\includegraphics[width=\textwidth]{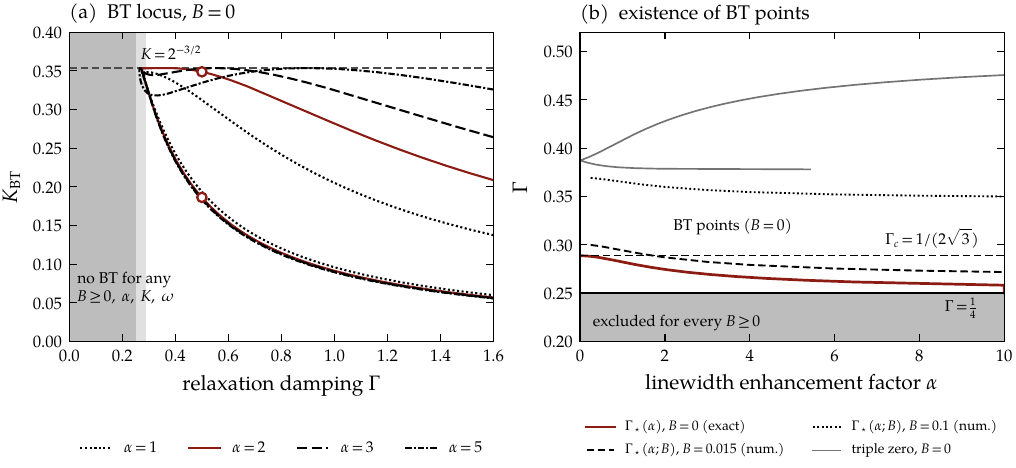}
\caption{Confinement of the BT degeneracy. (a)~BT injection amplitudes $K_{\mathrm{BT}}$ versus $\Gamma$ for $B=0$ and $\alpha=1,2,3,5$; each locus is born at $\Gamma_{\star}(\alpha)$ as a pair. Dark band: $\Gamma<\frac14$, excluded for every $B\ge0$ by Theorem~\ref{thm:barrier}; light band: $\frac14<\Gamma<\Gc$, where existence depends on $\alpha$. Dashed: the sum-rule bound $K\le2^{-3/2}$. Circles: the BT points of Sec.~\ref{sec:example}. (b)~Existence region in the $(\alpha,\Gamma)$ plane. Solid: $\Gamma_{\star}(\alpha)$ for $B=0$ from Theorem~\ref{thm:sharp}; dashed and dotted: numerical thresholds for $B=0.015$ and $B=0.1$; gray: the triple-zero curve \eqref{eq:TZ}, on which one BT point acquires $e_{1}=0$. Along the solid curve the BT points violate the transversality condition (G3) (Corollary~\ref{cor:loss}).}
\label{fig:confinement}
\end{figure}

Define $\Gamma_{\star}(\alpha)=\inf\{\Gamma>\frac14:\ \alpha\ge\alpha_{0}(\Gamma)\}$. Theorem~\ref{thm:sharp} gives $\frac14<\Gamma_{\star}(\alpha)\le\Gc$ for every $\alpha\ge0$ and $\Gamma_{\star}(\alpha)\to\frac14$ as $\alpha\to\infty$. The barrier \eqref{eq:quarter} is therefore optimal over the class of all linewidth enhancement factors, but not at any fixed one. Solving $\Phi_{\alpha}=\partial_{n}\Phi_{\alpha}=0$ numerically gives $\Gamma_{\star}(1)=0.28222$, $\Gamma_{\star}(2)=0.27455$, $\Gamma_{\star}(5)=0.26394$, $\Gamma_{\star}(20)=0.25464$ and $\Gamma_{\star}(100)=0.25109$, and these values increase toward $\Gc=0.28868$ as $\alpha$ decreases; for $\alpha=0$, item~(i) gives $\Gamma_{\star}(0)=\Gc$ exactly. At $\Gamma=\Gamma_{\star}(\alpha)$ two BT points are born together from a double root of $\Phi_{\alpha}$. On a grid of 119 values of $\Gamma$ in $(\frac14,\Gc)$ we found $\alpha_{0}$ strictly decreasing; we use this monotonicity only to draw Fig.~\ref{fig:confinement}(b) and do not claim it as a theorem. For $B>0$ we have no closed form; the thresholds computed numerically for $B=0.015$ and $B=0.1$ lie above $\Gamma_{\tan}(B)$, as they must.

%======================================================================
\section{Transversality of the unfolding by injection strength and detuning}\label{sec:transv}
%======================================================================

We now reduce condition (G3) to a property of the surface of locked states. Write $z=(x,y,n,K,\omega)\in\mathbb R^{5}$ and $f(z)$ for the right-hand side of \eqref{eq:model} in real coordinates. Near a BT point, $J(z)$ has one simple eigenvalue $\lambda_{3}(z)$ close to $\lambda_{3}\neq0$, a smooth function of $z$, and two eigenvalues $\mu_{1},\mu_{2}$ close to zero. Their sum and product,
\begin{equation}\label{eq:taudelta}
\tau(z):=\mu_{1}+\mu_{2}=e_{1}-\lambda_{3},\qquad \delta(z):=\mu_{1}\mu_{2}=\frac{e_{3}}{\lambda_{3}},
\end{equation}
are therefore smooth, and the characteristic polynomial factors as $(\lambda-\lambda_{3})(\lambda^{2}-\tau\lambda+\delta)$, so that
\begin{equation}\label{eq:e2e3taudelta}
e_{2}=\lambda_{3}\tau+\delta,\qquad e_{3}=\lambda_{3}\delta .
\end{equation}

We use an elementary lemma of linear algebra.

\begin{lemma}[Regular values on a submanifold]\label{lem:regular}
Let $F:U\subset\mathbb R^{m}\to\mathbb R^{k}$ and $h:U\to\mathbb R^{m-k}$ be smooth, $F(p)=0$ and $\rank DF(p)=k$, and let $\varphi$ be a local parametrization of $M=F^{-1}(0)$ near $p$, $\varphi(w_{0})=p$. For every $m\times k$ matrix $V$,
\begin{equation}\label{eq:lemreg}
\begin{aligned}
&\det\begin{pmatrix}DF(p)\\ Dh(p)\end{pmatrix}\cdot\det\bigl[D\varphi(w_{0})\ \ V\bigr]\\
&\qquad=(-1)^{k(m-k)}\det\bigl(DF(p)\,V\bigr)\,\det D(h\circ\varphi)(w_{0}).
\end{aligned}
\end{equation}
In particular the matrix on the left is invertible if and only if $h|_{M}$ is a local diffeomorphism at $p$.
\end{lemma}

\begin{proof}
Since $DF\,D\varphi=0$, the product of the $m\times m$ matrix $\binom{DF}{Dh}$ with $[D\varphi\ V]$ is block triangular, with diagonal blocks $DF\,V$ and $Dh\,D\varphi=D(h\circ\varphi)$; taking determinants gives \eqref{eq:lemreg}. Choosing $V$ with $\det(DF\,V)\neq0$, which is possible because $\rank DF=k$, and noting that $[D\varphi\ V]$ is then invertible, gives the last statement.
\end{proof}

\begin{proposition}[Transversality criterion]\label{prop:transv}
Let $p_{*}$ be a BT point of \eqref{eq:model} with $K>0$, and let
\begin{equation}\label{eq:C}
C:=\det\frac{\partial(e_{2},e_{3})}{\partial(n,s)}\bigg|_{(n_{*},s_{*})},
\end{equation}
where $e_{2}$ and $e_{3}$ are the functions \eqref{eq:e23} of the chart \eqref{eq:chart}, and write $D_{5}:=\det\partial(f,e_{2},e_{3})/\partial(x,y,n,K,\omega)$ and $D_{\tau\delta}:=\det\partial(f,\tau,\delta)/\partial(x,y,n,K,\omega)$. Then:
\begin{itemize}
\item[(i)] condition (G3) holds if and only if $C\neq0$, that is, if and only if the curves $e_{2}=0$ and $e_{3}=0$ cross transversally at $(n_{*},s_{*})$ in the plane of locked states;
\item[(ii)] at every locked state, $D_{5}=2\kappa K\det\partial(e_{2},e_{3})/\partial(n,s)$; in particular $D_{5}=2\kappa K\,C$ at $p_{*}$;
\item[(iii)] at $p_{*}$, $D_{5}=\lambda_{3}^{2}\,D_{\tau\delta}$;
\item[(iv)] if $B=0$, then $C=-(I/\alpha)\,\Phi_{\alpha}'(n_{*})$; hence (G3) holds if and only if $n_{*}$ is a simple root of $\Phi_{\alpha}$.
\end{itemize}
\end{proposition}

\begin{proof}
\emph{Rank of $Df$.} The $3\times3$ minor of $Df$ formed by the columns $x$, $y$ and $K$ is the first component of $J_{\cdot1}\times J_{\cdot2}$, which spans the left kernel of $J$ at a BT point. By \eqref{eq:J} and $E=-K/(\frac12n-is)$ it equals $\kappa(nx+2sy)=-2\kappa K\neq0$. Thus $\rank Df(p_{*})=3$ and, by the chart \eqref{eq:chart}, the zero set of $f$ near $p_{*}$ is the two-dimensional surface $M$ of locked states.

\emph{Reduction.} For $(K,\omega)$ near $(K_{*},\omega_{*})$ there is a two-dimensional parameter-dependent local center manifold, which contains every equilibrium close to $p_{*}$ and whose tangent plane at an equilibrium is invariant under $J$ \cite{Kuznetsov2004}. At an equilibrium on it, the eigenvalues of $D_{u}f_{c}$ are $\mu_{1},\mu_{2}$, so $\tr D_{u}f_{c}=\tau$ and $\det D_{u}f_{c}=\delta$. In coordinates $(u,w,K,\omega)$ in which the center manifold is $w=0$, the equation for $w$ has the form $\dot w=G$ with $G(u,0;K,\omega)\equiv0$ and $\partial_{w}G=\lambda_{3}\neq0$; hence $\rank Df=1+\rank D_{(u,K,\omega)}f_{c}$, and $D_{(u,K,\omega)}f_{c}$ has full rank $2$. The equilibria of $f_{c}$ form the same surface $M$, projected to the coordinates $(u,K,\omega)$. By Lemma~\ref{lem:regular}, applied once to $(f_{c},\tr,\det)$ and once to $(f,\tau,\delta)$, condition (G3) holds if and only if $(\tau,\delta)|_{M}$ is a local diffeomorphism at $p_{*}$, if and only if $\det\partial(f,\tau,\delta)/\partial z\neq0$.

\emph{From $(\tau,\delta)$ to $(e_{2},e_{3})$.} At $p_{*}$, $\tau=\delta=0$, so differentiating \eqref{eq:e2e3taudelta} gives $de_{2}=\lambda_{3}\,d\tau+d\delta$ and $de_{3}=\lambda_{3}\,d\delta$; the terms in $d\lambda_{3}$ drop out. The matrix $\bigl(\begin{smallmatrix}\lambda_{3}&1\\0&\lambda_{3}\end{smallmatrix}\bigr)$ has determinant $\lambda_{3}^{2}\neq0$, which proves (iii) and shows that (G3) holds if and only if $(e_{2},e_{3})|_{M}$ is a local diffeomorphism. In the chart $(n,s)$ this is $C\neq0$, which proves (i).

\emph{Formula (ii).} Apply \eqref{eq:lemreg} with $m=5$, $k=3$, $F=f$, $h=(e_{2},e_{3})$, $\varphi$ the chart \eqref{eq:chart} and $V=[\,\mathbf e_{x}\ \mathbf e_{y}\ \mathbf e_{K}\,]$. Then $\det(Df\,V)=-2\kappa K$ (computed above, valid at every locked state), and $\det[D\varphi\ V]=-1$, because the rows $n$ and $\omega$ of $D\varphi=[\partial_{n}\varphi\ \partial_{s}\varphi]$ are $(1,0)$ and $(\alpha/2,1)$. Hence $D_{5}=2\kappa K\det\partial(e_{2},e_{3})/\partial(n,s)$, which equals $2\kappa K\,C$ at $p_{*}$.

\emph{Formula (iv).} For $B=0$, by \eqref{eq:e3B0}, at a point with $e_{2}=0$ and $\alpha s=g$ the gradient of $e_{3}$ is $I(g',-\alpha)-2\Gamma\nabla e_{2}$, while $\nabla e_{2}=(-\sigma',2s)$. Therefore $C=\det\bigl(\begin{smallmatrix}-\sigma'&2s\\ Ig'&-\alpha I\end{smallmatrix}\bigr)=I(\alpha\sigma'-2sg')=-(I/\alpha)(2gg'-\alpha^{2}\sigma')=-(I/\alpha)\Phi_{\alpha}'$.
\end{proof}

Proposition~\ref{prop:transv} separates the four ingredients of genericity. Nondegeneracy of the germ, (G1)--(G2), is a property of the vector field at the point; hyperbolicity of the transverse eigenvalue, $\lambda_{3}\neq0$, is part of (G0); transversality (G3) is a property of how $(K,\omega)$ move the equilibrium; and genericity of the codimension-two point is the conjunction of the four. The criterion (i) involves only the closed forms \eqref{eq:e23} and no eigenvector. Item (iv) has an immediate consequence at the boundary of the existence region.

\begin{corollary}[Loss of transversality at the existence threshold]\label{cor:loss}
Let $B=0$, $\Gamma>\frac14$ and $\alpha=\alpha_{0}(\Gamma)>0$. At every point $n_{0}$ where the infimum \eqref{eq:alpha0} is attained and $e_{1}\neq0$, the locked state is a BT point at which (G3) fails.
\end{corollary}

\begin{proof}
For $\alpha=\alpha_{0}(\Gamma)$, $\Phi_{\alpha}=\sigma\,(g^{2}/\sigma-\alpha^{2})\ge0$ on the interior of $W_{\Gamma}$, with equality at $n_{0}$, which is interior because the quotient in \eqref{eq:alpha0} tends to $+\infty$ at both ends when $\alpha_{0}>0$. Thus $n_{0}$ is a root of $\Phi_{\alpha}$ and an interior minimum, so $\Phi_{\alpha}'(n_{0})=0$ and $C=0$ by Proposition~\ref{prop:transv}(iv).
\end{proof}

The curve $\Gamma_{\star}(\alpha)$ of Fig.~\ref{fig:confinement}(b) was computed from $\Phi_{\alpha}=\Phi_{\alpha}'=0$. At each of its points the BT conditions have a double root, so that, by Proposition~\ref{prop:transv}(iv), the degeneracy exists but is not a generic codimension-two point of the $(\omega,K)$ plane: it is the point where two BT points are born. Numerically, for $\alpha=2$ the two BT points that exist for $\Gamma$ slightly above $\Gamma_{\star}(2)$ have values of $C$ of opposite sign that vanish like $(\Gamma-\Gamma_{\star})^{1/2}$. For $\alpha>\alpha_{0}(\Gamma)$, every BT point at a simple root of $\Phi_{\alpha}$ satisfies (G3).

%======================================================================
\section{Contacts between Hopf and saddle-node curves}\label{sec:contacts}
%======================================================================

\subsection{Exact curves}

Because every locked state is labeled by $(n,s)$ and the injection parameters are recovered from \eqref{eq:chart}, each bifurcation curve of locked states in the $(\omega,K)$ plane is the image of a curve in the $(n,s)$ plane. Both relevant conditions are quadratic in $s$.

\begin{proposition}[Parametrizations]\label{prop:curves}
Let $\xi=-2\Gamma/\kappa$ and $I=1-2\Gamma n/\kappa$. The saddle-node curve $\mathrm{SN}=\{e_{3}=0\}$ and the curve $\{e_{1}e_{2}=e_{3}\}$ are the images under \eqref{eq:chart} of
\begin{gather}
\xi s^{2}-\alpha\kappa I\,s+\xi\frac{n^{2}}{4}+\kappa I\frac n2=0,\label{eq:SNcurve}\\
n s^{2}+\alpha\kappa I\,s+\frac{n^{3}}{4}+(n+\xi)\bigl[(1+\kappa)I-1\bigr]-\kappa I\frac n2=0.\label{eq:Hcurve}
\end{gather}
The second curve is the Hopf curve $\mathrm{H}$ where $e_{2}>0$, with frequency $e_{2}^{1/2}$, and a neutral-saddle curve where $e_{2}<0$.
\end{proposition}

\begin{proof}
Equation~\eqref{eq:SNcurve} is $e_{3}=0$ in the form \eqref{eq:e23}; \eqref{eq:Hcurve} is $e_{1}e_{2}-e_{3}=0$ after the cancellation $(n+\xi)d-\xi d=nd$. For a cubic $\lambda^{3}-e_{1}\lambda^{2}+e_{2}\lambda-e_{3}$, a pair $\pm i\nu$ occurs if and only if $e_{1}e_{2}=e_{3}$ and $\nu^{2}=e_{2}>0$.
\end{proof}

These are the analytic stability boundaries of Ref.~\cite{Gavrielides1997}, written in the variables of the gain--loss frame.

\subsection{Classification of contacts}

\begin{corollary}[Contacts between Hopf and saddle-node]\label{cor:contact}
Let a locked state belong to the closure $\overline{\mathrm H}$ of the Hopf curve and to $\mathrm{SN}$. Then $e_{1}e_{2}=0$ and $e_{2}\ge0$, and exactly one of the following holds:
\begin{itemize}
\item[(a)] $e_{1}=0<e_{2}$: a fold--Hopf (zero--Hopf, ZH) point, with spectrum $\{0,\pm i\sqrt{e_{2}}\}$;
\item[(b)] $e_{2}=0\neq e_{1}$: a BT point in the sense of Definition~\ref{def:BT};
\item[(c)] $e_{1}=e_{2}=0$: a triple-zero point, at which $J$ is nilpotent with a single Jordan block of size three.
\end{itemize}
Moreover: if $\Gamma<\frac14$ (any $B\ge0$), only (a) occurs; if $B=0$ and $\Gamma\ge\frac12$, only (b) occurs; and if $B=0$, case (c) occurs exactly on the curve \eqref{eq:TZ}.
\end{corollary}

\begin{proof}
On $\mathrm{SN}$, $e_{3}=0$, so $e_{1}e_{2}=e_{3}$ gives $e_{1}e_{2}=0$; $e_{2}\ge0$ on $\overline{\mathrm H}$ by continuity. The three cases are exhaustive and mutually exclusive. In (a) the characteristic polynomial is $\lambda(\lambda^{2}+e_{2})$. In (c) it is $\lambda^{3}$, so $J$ is nilpotent; since $\rank J\ge2$ at every locked state (Sec.~\ref{sec:BT}), $\rank J=2$, and a nilpotent $3\times3$ matrix of rank two consists of a single Jordan block. If $\Gamma<\frac14$, the proof of Theorem~\ref{thm:barrier} gives $e_{2}>0$ at every locked state for every $B\ge0$, which excludes (b) and (c). If $B=0$, $e_{1}=n-2\Gamma$ vanishes only at $n=2\Gamma$, where $I=1-4\Gamma^{2}\le0$ when $\Gamma\ge\frac12$; this excludes (a) and (c). The last statement is the final part of Theorem~\ref{thm:sharp}.
\end{proof}

For class-B parameters this is the familiar picture of two saddle-node--Hopf points on the locking boundary \cite{Wieczorek2005}; we find two such points for every $\alpha$ tested in Corollary~\ref{cor:classB}, and Corollary~\ref{cor:contact} explains why they cannot be BT points. Corollary~\ref{cor:contact} classifies the \emph{linear} type of a contact. It does not assert that a triple-zero point is a generic codimension-three singularity: the unfolding of the nilpotent triple-zero singularity requires nondegeneracy conditions on higher-order terms \cite{Dumortier2001}, which we have not verified. In the $(\omega,K)$ plane at fixed $(\Gamma,\alpha,B)$, triple-zero points occur only on a curve in the $(\Gamma,\alpha)$ plane, so they are not encountered in a generic two-parameter diagram; they mark the values of $(\Gamma,\alpha)$ at which the type of contact changes.

For $B=0$ and $\alpha=2$ the triple-zero curve \eqref{eq:TZ} is crossed at $\Gamma_{\mathrm{TZ},1}=0.378621437658769$ and $\Gamma_{\mathrm{TZ},2}=0.427993484037456$. At both values the three curves $e_{1}=0$, $e_{2}=0$, $e_{3}=0$ cross pairwise transversally in the $(n,s)$ plane, and a ZH point and a BT point coalesce: the point of SN with $e_{1}=0$ has $e_{2}>0$ (a ZH point) for $\Gamma$ below the triple-zero value and $e_{2}<0$ above it, where it is no longer a contact with the Hopf curve, while the transverse eigenvalue $e_{1}$ of the BT point changes from positive to negative. Accordingly, for $\alpha=2$ we find two ZH points at $\Gamma=0.035$, $\frac14$, $0.30$ and $0.37$, one at $\Gamma=0.38$ and $0.40$, and none at $\Gamma=0.43$, $0.45$ and $\frac12$. Both BT points have $e_{1}>0$ for $\Gamma_{\star}(2)<\Gamma<\Gamma_{\mathrm{TZ},1}$, so that the transverse direction is repelling; one of them has $e_{1}<0$ between the two triple-zero values, and both have $e_{1}<0$ above $\Gamma_{\mathrm{TZ},2}$. For $B=0$ the first statement holds for every $\alpha$ when $\Gamma<1/\sqrt7$: every BT point has $n\ge n_{-}$, and $n_{-}>2\Gamma$ is equivalent to $\Gamma^{2}<1/7$, so that $e_{1}=n-2\Gamma>0$. Beyond that, $e_{1}$ can change sign along a branch of BT points only on the triple-zero curve.

%======================================================================
\section{Bifurcation geometry and periodic orbits}\label{sec:geometry}
%======================================================================

\subsection{Diagrams}

\begin{figure}[t]
\centering
\includegraphics[width=\textwidth]{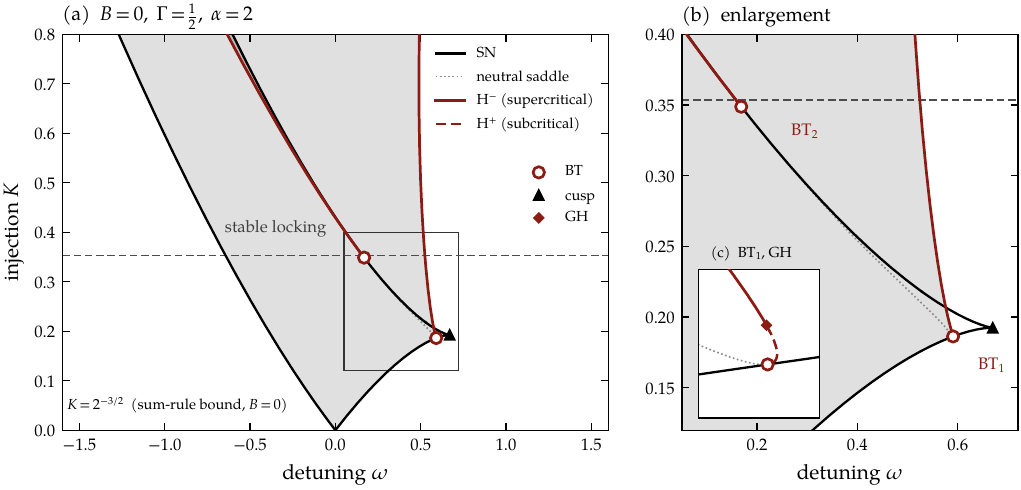}
\caption{Bifurcation diagram of Eq.~\eqref{eq:model} in the plane of detuning $\omega$ and injection $K$ for $B=0$, $\Gamma=\frac12$, $\alpha=2$. Light (dark) gray: one (two) stable locked states. Black: saddle-node curve SN with its cusp (triangle). Dark red: Hopf curve, supercritical (solid, $\ell_{1}<0$) and subcritical (dashed, $\ell_{1}>0$); dotted: neutral saddle. Open circles: $\mathrm{BT}_{1}$ and $\mathrm{BT}_{2}$; diamond: generalized Hopf point GH. Dashed horizontal: the sum-rule bound $K=2^{-3/2}$. (b)~Enlargement of the box in (a); (c)~enlargement around $\mathrm{BT}_{1}$ showing the change of criticality at GH.}
\label{fig:main}
\end{figure}

Figure~\ref{fig:main} shows the bifurcation diagram at $B=0$, $\Gamma=\frac12$, $\alpha=2$. Shading counts stable locked states (one or two), obtained by solving the cubic locking equation on a grid and applying the Routh--Hurwitz conditions $e_{1}<0$, $e_{2}>0$, $e_{3}<0$, $e_{1}e_{2}<e_{3}$. All curves are drawn from Proposition~\ref{prop:curves}; no numerical continuation of equilibria is involved. Criticality of the Hopf curve is given by the sign of the first Lyapunov coefficient $\ell_{1}$, computed with the projection formula of Ref.~\cite[Sec.~3.5]{Kuznetsov2004}. The saddle-node curve has one cusp, at $(\omega,K)=(0.669873,0.192414)$, where the cubic $\kappa F$ has a triple root. The two BT points sit on the saddle-node curve, below the sum-rule bound $K=2^{-3/2}$, and are joined by a neutral-saddle segment. Close to $\mathrm{BT}_{1}$ the Hopf curve changes criticality at a generalized Hopf point, $(\omega,K)=(0.5907493,0.1865616)$, where $\ell_{1}=0$ with Hopf frequency $0.163$. The sign of $\ell_{1}$ on the Hopf curve next to each BT point agrees with the sign of $ab$ of the BT normal form (Sec.~\ref{sec:example}), which is an independent consistency test.

\begin{figure}[t]
\centering
\includegraphics[width=\textwidth]{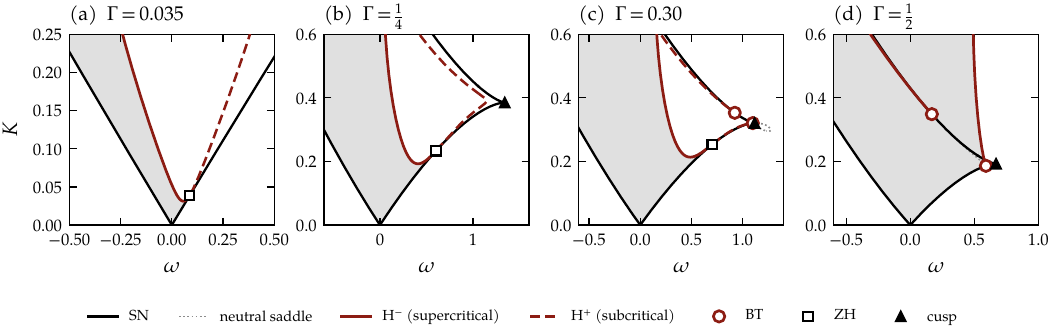}
\caption{The same diagram for $\alpha=2$, $B=0$ and increasing damping: (a)~$\Gamma=0.035$, (b)~$\Gamma=\frac14$, (c)~$\Gamma=0.30$, (d)~$\Gamma=\frac12$. Squares: fold--Hopf (ZH) points. Below $\Gamma_{\star}(2)=0.27455$ every contact between H and SN is a ZH point; at $\Gamma=\frac12$ every contact is a BT point. Windows differ between panels.}
\label{fig:sequence}
\end{figure}

Figure~\ref{fig:sequence} follows the same diagram as the damping is increased at fixed $\alpha=2$, $B=0$. For $\Gamma=0.035$ and $\Gamma=\frac14$ the Hopf curve meets the saddle-node curve only at ZH points, where it changes criticality; at $\Gamma=\frac14$ the subcritical branch runs close to the saddle-node near the cusp. At $\Gamma=0.30>\Gamma_{\star}(2)$ two BT points, both with $e_{1}>0$, coexist with a ZH point; a second ZH point lies outside the window shown, at $(\omega,K)\approx(-1.64,1.80)$. At $\Gamma=\frac12$ only BT contacts remain, in accordance with Corollary~\ref{cor:contact}. The transition between the two kinds of contact passes through the triple-zero points described in Sec.~\ref{sec:contacts}.

\subsection{Spectral signature along the saddle-node}

\begin{figure}[t]
\centering
\includegraphics[width=\textwidth]{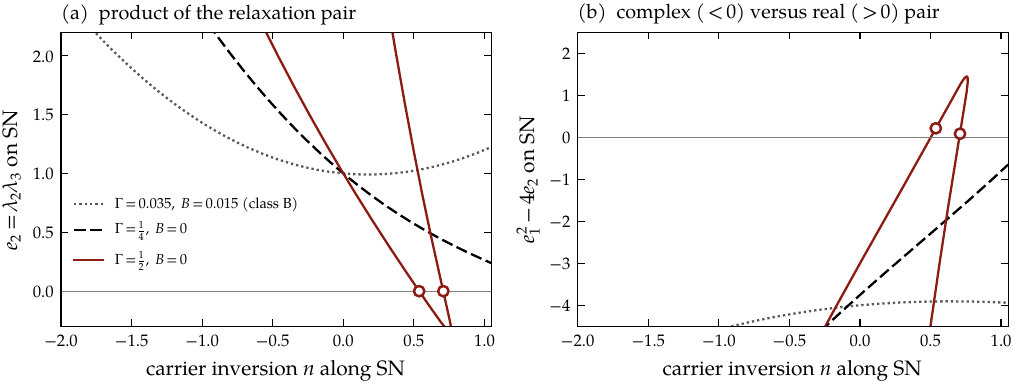}
\caption{Spectral signature of the relaxation pair along the saddle-node curve, parametrized by the inversion $n$, for $\alpha=2$. (a)~$e_{2}$, the product of the two nonzero eigenvalues; (b)~$e_{1}^{2}-4e_{2}$, negative for a complex pair. Dotted: $\Gamma=0.035$, $B=0.015$; dashed: $\Gamma=\frac14$, $B=0$; solid: $\Gamma=\frac12$, $B=0$. Circles: BT points. All curves pass through $e_{2}=1$ at $n=0$, the free-running limit.}
\label{fig:signature}
\end{figure}

Figure~\ref{fig:signature} displays the quantity that Theorem~\ref{thm:barrier} controls. Along SN, $e_{2}$ starts at one in the weak-injection limit and stays close to one for class-B parameters, whereas for $\Gamma=\frac12$ it crosses zero at the two BT points. Panel (b) shows the discriminant $e_{1}^{2}-4e_{2}$ of the pair: at a BT point the pair is real, $\{0,e_{1}\}$, so that along any path on SN from weak injection to a BT point the relaxation pair must turn from complex to real when $\Gamma<1$. For class-B parameters it remains complex on the part of the saddle-node curve with $0<n<2$, the only range in which Theorem~\ref{thm:sum} allows BT points.

\subsection{Periodic orbits near \texorpdfstring{$\mathrm{BT}_{2}$}{BT2}}\label{sec:cycles}

\begin{figure}[t]
\centering
\includegraphics[width=\textwidth]{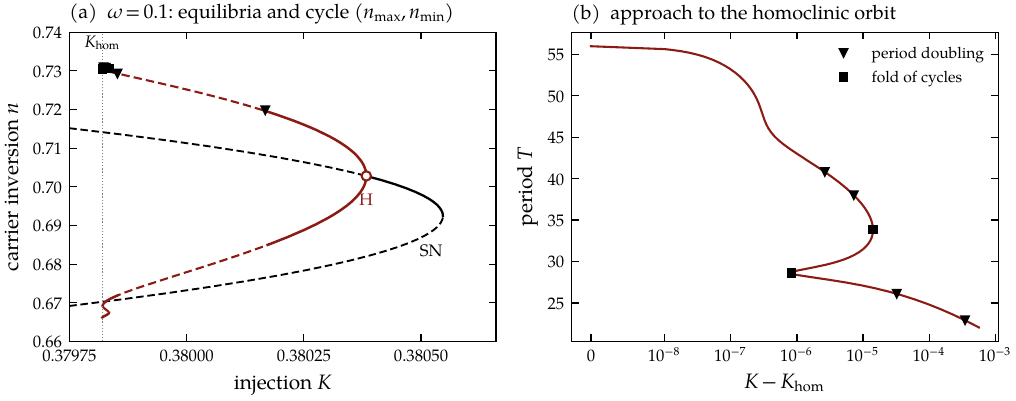}
\caption{Periodic orbits near $\mathrm{BT}_{2}$ along $\omega=0.1$ ($B=0$, $\Gamma=\frac12$, $\alpha=2$). (a)~Locked states (black; solid stable, dashed unstable; the stable low-inversion state $n\approx-0.22$ lies below the window) and the periodic orbit born at the supercritical Hopf point H, shown by the extrema of $n$ (dark red; solid stable, dashed unstable). Triangles: period doubling; squares: folds of cycles; dotted line: the homoclinic value $K_{\mathrm{hom}}$. (b)~Period along the branch versus $K-K_{\mathrm{hom}}$ (logarithmic scale), showing the approach to the homoclinic orbit.}
\label{fig:cycles}
\end{figure}

To follow the unfolding of $\mathrm{BT}_{2}$ beyond the local curves, we continued the periodic orbit born at the Hopf point along the slice $\omega=0.1$ (Fig.~\ref{fig:cycles}) by single shooting with pseudo-arclength continuation; Floquet multipliers were obtained from the variational equations. Two independent implementations were used (Wolfram Language and Python/SciPy, with relative tolerances $10^{-11}$ and $10^{-12}$), and the values quoted below agree between them to the digits given. On this slice the Hopf point is at $K_{\mathrm H}=0.3803832$ and the saddle-node at $K_{\mathrm{SN}}=0.3805466$. The Hopf bifurcation is supercritical, as expected from $\sign(ab)=-1$ at $\mathrm{BT}_{2}$, and the orbit born there is stable; a direct simulation at $K=0.380205$ converges to it with period $22.72645$. As $K$ decreases the orbit grows toward the saddle locked state and its period increases without bound; the branch converges to $K_{\mathrm{hom}}\simeq0.379820$, where the largest computed period exceeds $55$. We located the same value independently, without continuation, from the change in the fate of the unstable manifold of the saddle: for $K$ slightly below $K_{\mathrm{hom}}$ both branches of the one-dimensional unstable manifold end at the stable low-inversion locked state, while for $K$ slightly above one branch does not. Bisection of this criterion gives $K_{\mathrm{hom}}=0.37982017$. Both computations are numerical evidence of a homoclinic orbit to the saddle at $K_{\mathrm{hom}}$; we do not prove its existence.

At $K_{\mathrm{hom}}$ the saddle is a saddle-focus, with unstable eigenvalue $\lambda_{u}=0.2230$ and stable pair of real part $-0.2764$, so that the saddle quantity $\lambda_{u}+\operatorname{Re}\lambda_{s}=-0.053$ is negative. For a negative saddle quantity, Shil'nikov's theorem gives a unique stable periodic orbit close to a homoclinic loop that is broken transversally by the parameter \cite[Chap.~6]{Kuznetsov2004}, and no chaotic set near the loop; in the computation the branch is monotone and stable for periods above $41$. The hypotheses of the theorem on the global behavior of the loop were not verified, and the statement is used only to interpret the computation. Farther from $K_{\mathrm{hom}}$ the orbit loses and regains stability at four period-doubling points ($K=0.3801698$, $0.3798518$, $0.3798275$, $0.3798229$) and passes two folds of cycles ($K=0.3798344$, $0.3798211$); these events lie outside the neighborhood described by the theorem. Period-doubled branches were not continued; a direct simulation at $K=0.3800$, between the first two period doublings, settles on an attractor that is not the period-one orbit (its successive maxima of $n$ take many distinct values), and we do not characterize it here. The homoclinic curve lies in the thin wedge between H and SN that emanates from $\mathrm{BT}_{2}$, of width $7\times10^{-4}$ in $K$ at $\omega=0.1$, and is not resolved at the scale of Fig.~\ref{fig:main}.

%======================================================================
\section{Two generic Bogdanov--Takens points constructed from the physics}\label{sec:example}
%======================================================================

We now build BT points by following the logic of the previous sections rather than by solving the characteristic polynomial, and verify the four conditions of genericity.

\emph{Step 1: fix the device.} Take $B=0$, $\Gamma=\frac12$, $\alpha=2$. Since $\Gamma>\Gc$, Theorem~\ref{thm:sharp} guarantees BT points.

\emph{Step 2: the balance fixes $\xi$.} By Lemma~\ref{lem:xi}, $\xi=-2\Gamma=-1$ exactly, from the balance law alone.

\emph{Step 3: the spectral conditions fix the state.} With $\xi=-1$, $I=1-n$, and Eqs.~\eqref{eq:BTaffine} become two equations in $(n,s)$, or $\Phi_{2}(n)=0$ on $W_{1/2}=[4-2\sqrt3,1)$, with $4-2\sqrt3\approx0.5359$. There are exactly two roots (certified in Step~6), and the injection parameters follow from \eqref{eq:chart}. Their values are listed in Table~\ref{tab:BT}.

\emph{Step 4: spectral verification.} On the reconstructed field the residual of the vector field is below $10^{-50}$ at both points; the singular values of $J$ are $(1.78412,\,0.628477)$ and a third one below $10^{-50}$ at $\mathrm{BT}_{1}$, and $(1.71572,\,0.648179)$ and a third one below $10^{-50}$ at $\mathrm{BT}_{2}$, so $\rank J=2$ and the zero eigenvalue has geometric multiplicity one. With $\xi=-1$ the transverse eigenvalue $e_{1}=n-1$ equals $-I_{*}$ exactly, so (G0) holds with $\lambda_{3}<0$ and both center manifolds attract. The sum rule holds to working precision and both points satisfy \eqref{eq:bounds}. The reduced data give $-\nn_{3}/\dd_{2}=-\nn_{2}/\dd_{1}=-1$; as explained in Sec.~\ref{sec:BT}, this agreement is forced by \eqref{eq:BTaffine} once the point is a BT point, and the physics resides in the fact that $-1$ is the value prescribed by the balance law.

\emph{Step 5: the germ, (G1)--(G2).} With $q_{0}$ the unit vector spanning $\ker J$, oriented so that its $y$-component is negative, and the Jordan chains of \ref{app:normal}, the quadratic coefficients of the BT normal form are listed in Table~\ref{tab:BT}. Rescaling $q_{0}$ by $c$ multiplies $(a,b)$ by $c$, so the invariant statements are $ab\neq0$ and $\sign(ab)=+1$ at $\mathrm{BT}_{1}$, $-1$ at $\mathrm{BT}_{2}$. The two BT points are of opposite type: the Hopf curve emanating from $\mathrm{BT}_{1}$ is subcritical and the one from $\mathrm{BT}_{2}$ supercritical, as confirmed by $\ell_{1}$ in Fig.~\ref{fig:main}. The small ratio $|b/a|$ at $\mathrm{BT}_{1}$ is consistent with the generalized Hopf point found close to it.

\emph{Step 6: the unfolding, (G3).} By Proposition~\ref{prop:transv}(iv), (G3) holds if and only if $\Phi_{2}'(n_{*})\neq0$. For $\Gamma=\frac12$, $\alpha=2$, one has $\Phi_{2}=P(n)/[4(1-n)^{2}]$ with $P$ a quartic polynomial with rational coefficients. An exact Sturm count shows that $P$ has exactly two roots, counted with multiplicity, in $[4-2\sqrt3,1]$. For each of them, Table~\ref{tab:cert} gives an isolating interval $[n_{\mathrm{lo}},n_{\mathrm{hi}}]$ of width $10^{-20}$ with rational endpoints, the signs of $\Phi_{2}$ at the endpoints, which are opposite, and an enclosure of $\Phi_{2}'$ on the interval obtained by exact interval arithmetic, which excludes zero. Hence each interval contains exactly one root and the root is simple: this is a computer-assisted proof that the unfolding by $(K,\omega)$ is transversal at both points. The same intervals certify (G0): at every root of $\Phi_{2}$, $s=g/2$ gives $e_{2}=e_{3}=0$ exactly by \eqref{eq:e3B0}, while $e_{1}=n-1\in[n_{\mathrm{lo}}-1,n_{\mathrm{hi}}-1]\subset(-1,0)$ and $I=1-n>0$. As a check, the three determinants of Proposition~\ref{prop:transv} were computed independently from the $5\times5$ Jacobians, with $\tau$ and $\delta$ differentiated through first-order perturbation of the simple eigenvalue $\lambda_{3}$; the identities (ii)--(iv) hold to the working precision.

\begin{table}[t]
\centering
\caption{The two BT points at $B=0$, $\Gamma=\frac12$, $\alpha=2$. All quantities were computed with 60-digit arithmetic from the defining equations; coordinates are given to 17 decimal places and the remaining quantities to 15 significant digits. $C$, $D_{5}$ and $D_{\tau\delta}$ are the determinants defined in Proposition~\ref{prop:transv}. The interval certificate for (G0) and (G3) is given in Table~\ref{tab:cert}.}
\label{tab:BT}
\vspace{4pt}
\small
\begin{tabular}{@{}lll@{}}
\toprule
 & $\mathrm{BT}_{1}$ & $\mathrm{BT}_{2}$\\
\midrule
$n_{*}$ & $0.53753419530643666$ & $0.70893033878599020$\\
$s_{*}$ & $0.05322253124324996$ & $-0.54056925667223778$\\
$\omega_{*}$ & $0.59075672654968662$ & $0.16836108211375242$\\
$K_{*}$ & $0.18632381402234443$ & $0.34875000481079985$\\
$I_{*}$ & $0.46246580469356334$ & $0.29106966121400980$\\
$\lambda_{3}=e_{1}=-I_{*}$ & $-0.46246580469356334$ & $-0.29106966121400980$\\
\midrule
$a$ (G1) & $-0.304779098708819$ & $-2.69215215003489$\\
$b$ (G2) & $-0.0290755350343872$ & $19.9160331083730$\\
$\sign(ab)$ & $+1$ (subcritical) & $-1$ (supercritical)\\
\midrule
$C$ (G3) & $1.80682693800157$ & $-2.59908961627203$\\
$\Phi_{2}'(n_{*})$ & $-7.81388340354723$ & $17.8588837148578$\\
$D_{5}=2\kappa K C$ & $0.673309772733535$ & $-1.81286503235714$\\
$D_{\tau\delta}=D_{5}/\lambda_{3}^{2}$ & $3.14815180560112$ & $-21.3979215679672$\\
\bottomrule
\end{tabular}
\end{table}

\begin{proposition}[Genericity of $\mathrm{BT}_{1}$ and $\mathrm{BT}_{2}$]\label{prop:generic}
For $B=0$, $\Gamma=\frac12$, $\alpha=2$, the two locked states of Table~\ref{tab:BT} satisfy (G0) and (G3), with computer assistance (exact Sturm count and exact interval arithmetic, Table~\ref{tab:cert}). The coefficients $a$ and $b$, evaluated in 60-digit floating-point arithmetic, are far from zero relative to the working precision; on this numerical evidence (G1)--(G2) hold as well, and the two points are generic codimension-two BT points of the $(\omega,K)$ unfolding, with a subcritical Hopf curve emanating from $\mathrm{BT}_{1}$ and a supercritical one from $\mathrm{BT}_{2}$.
\end{proposition}

The proposition distinguishes the status of its parts: (G0) and (G3) are established with computer assistance; (G1)--(G2) rest on high-precision floating-point evaluation of $a$ and $b$, which we did not enclose rigorously.

%======================================================================
\section{Physical interpretation, predictions and open questions}\label{sec:physics}
%======================================================================

We separate five levels.

\emph{Exact results of the model.} Lemmas~\ref{lem:xi}, \ref{lem:fold} and \ref{lem:regular}, Propositions~\ref{prop:closed}, \ref{prop:tan}, \ref{prop:transv} and \ref{prop:curves}, Theorems~\ref{thm:sum}, \ref{thm:barrier} and \ref{thm:sharp}, and Corollaries~\ref{cor:classB}, \ref{cor:loss} and \ref{cor:contact} are theorems about Eq.~\eqref{eq:model} and hold for all parameter values stated. Proposition~\ref{prop:generic} is a statement about two specific points: its conditions (G0) and (G3) are established with computer assistance, and (G1)--(G2) rest on high-precision numerical evaluation. The numerical values of $\Gamma_{\star}$, $\Gamma_{\tan}$, the class-B minima, the cusp, the generalized Hopf point and the periodic orbits are numerical results.

\emph{Interpretation.} In the units of the model the free-running relaxation pair has frequency one and damping $\Gamma$. The barrier says that the BT degeneracy needs a relaxation oscillation whose damping rate is at least one quarter of its frequency. By \eqref{eq:scaling}, condition \eqref{eq:quarter} reads
\begin{equation}\label{eq:quarterphys}
\begin{gathered}
\frac{\gamma_{N}}{\omega_{R}}+\frac{\omega_{R}}{\gamma_{E}}\ \ge\ \frac12,
\qquad\text{equivalently}\qquad
\omega_{R}\le\omega_{-}\ \ \text{or}\ \ \omega_{R}\ge\omega_{+},\\[2pt]
\omega_{\pm}=\frac{\gamma_{E}}{4}\Bigl[1\pm\Bigl(1-\frac{16\gamma_{N}}{\gamma_{E}}\Bigr)^{1/2}\Bigr],
\end{gathered}
\end{equation}
where the second form applies when $\gamma_{N}/\gamma_{E}<1/16$; otherwise \eqref{eq:quarter} holds for every $\omega_{R}$. In the class-B regime $\gamma_{N}\ll\gamma_{E}$ one has $\omega_{-}\simeq2\gamma_{N}$ and $\omega_{+}\simeq\gamma_{E}/2$: within the model, the degeneracy requires either a relaxation-oscillation frequency not larger than about twice the carrier decay rate, which happens close to threshold since $\omega_{R}$ grows with the pump \cite{ErneuxGlorieux2010}, or a relaxation-oscillation frequency comparable to half the photon decay rate. For the class-B values of Ref.~\cite{Wieczorek2005} the left-hand side of \eqref{eq:quarterphys} is $0.07$. The minimum of $\Gamma$ over $\omega_{R}$ is $(\gamma_{N}/\gamma_{E})^{1/2}$, attained at $\omega_{R}=(\gamma_{N}\gamma_{E})^{1/2}$.

The distinction with the familiar notion of damping is worth stating. The damping rate of the relaxation oscillation, $\Gamma\omega_{R}$, is a property of the free-running laser; the barrier is a statement about the ratio $\Gamma$ of this rate to the oscillation frequency, and it controls a spectral quantity of the injected laser, the product $e_{2}$ of the two eigenvalues that remain at a saddle-node. The transport rate $\xi=-2\Gamma/\kappa$ is the bridge: it is the rate at which carriers return to the balance at fixed field, it appears as the only nontrivial coupling of the inversion to itself in the spectral problem, and for $B=0$ it equals $-2\Gamma$ exactly. The quantity that permits or forbids the degeneracy is therefore the strength of carrier relaxation relative to the relaxation-oscillation frequency, not the injection or the detuning, which enter the bound only through $s^{2}$.

\emph{Testable predictions.} Within Eq.~\eqref{eq:model}: (i) whenever $\Gamma<\frac14$, measured as the damping-to-frequency ratio of the free-running relaxation oscillation, Hopf and saddle-node boundaries of the locking region can meet only at fold--Hopf points, at which the oscillation that appears has a nonzero frequency, at least $\omega_{R}(1-16\Gamma^{2})^{1/2}$ in physical units; (ii) on every saddle-node boundary, the product $e_{2}$ of the two nonzero eigenvalues of the locked laser is bounded below by $1-16\Gamma^{2}$ in the units of the model; where these two eigenvalues form a complex-conjugate pair, their common modulus is $|\lambda|=e_{2}^{1/2}$, which is therefore at least $\omega_{R}(1-16\Gamma^{2})^{1/2}$ in physical units. Where the pair is real, as near a BT point, the bound constrains only the product of the two eigenvalues and not each of them. In the complex-pair regime both quantities are accessible in principle through the relaxation-oscillation peak of the intensity-noise or optical spectra of the free-running and of the locked laser. (iii) If a device can be operated with $\Gamma$ above the threshold $\Gamma_{\star}(\alpha)$ of Sec.~\ref{sec:sharp}, the transition between fold--Hopf and BT contacts should be preceded by a slowing down of the oscillation at the locking boundary, since at a BT point the Hopf frequency $e_{2}^{1/2}$ tends to zero.

\emph{Existing evidence.} The quantitative agreement between bifurcation diagrams of Eq.~\eqref{eq:model} and measured stability maps \cite{Wieczorek2002,Wieczorek2003} was obtained in the class-B regime, where the locking boundary is organized by saddle-node--Hopf points; this is consistent with the barrier, but it is not a test of it, because no experiment we know of has explored the region $\Gamma\ge\frac14$ in the same configuration. The phase-locking dynamics of weakly damped devices, such as quantum-well lasers, and of highly damped devices are known to differ qualitatively \cite{Kelleher2012}, and excitability at the locking boundary has been observed in a quantum-dot laser \cite{Goulding2007}. We make no claim about the values of $\Gamma$ in these experiments.

\emph{Limitations and open questions.} The model has three variables and no spontaneous-emission noise, thermal effects, gain saturation in the sense of a compression coefficient, spatial or multimode dynamics, or carrier-dependent coupling beyond the linewidth enhancement factor. Whether a given device can be operated, within the validity of Eq.~\eqref{eq:model}, at $\Gamma\ge\frac14$ is a question about specific lasers that we do not settle. For $B>0$ the exact threshold is known only numerically; a closed-form extension of Theorem~\ref{thm:sharp} and a proof of the monotonicity of $\alpha_{0}$ are open. The nondegeneracy of the triple-zero points and the global properties of the homoclinic orbit near $\mathrm{BT}_{2}$ remain to be established.

%======================================================================
\section{Conclusions}\label{sec:conclusions}
%======================================================================

A Bogdanov--Takens point is usually located; here we have asked which property of the physical balance of the device permits or forbids it. For the optically injected semiconductor laser the answer is written in the carrier balance. The scalar that couples the reduced spectrum to the full one is the slope of the gain--loss balance, $\xi=-2\Gamma/\kappa$, a physical rate before it is a spectral quantity. In that frame the two BT conditions become affine and yield the sum rule $K^{2}=I[1-(1+\kappa)I]$, which confines BT points to depleted intensity, moderate injection and $0<n<2$. The relaxation oscillation then supplies the barrier: the product of the two nonzero eigenvalues of every saddle-node is bounded below by $1-16\Gamma^{2}$, so that for $\Gamma<\frac14$ the relaxation pair cannot be brought down to the phase mode, for any $B$, linewidth enhancement factor, injection strength or detuning. For $B=0$ the confinement is exact, with threshold $\Gamma_{\star}(\alpha)$ between $\frac14$ and $1/(2\sqrt3)$.

The same frame decides whether an existing BT point organizes the standard unfolding in the plane of injection strength and detuning. Transversality of the unfolding is equivalent to the transversal crossing of the curves $e_{2}=0$ and $e_{3}=0$ in the plane of locked states, a condition that involves only closed forms. It holds at simple roots of $\Phi_{\alpha}$ and fails exactly on the existence threshold, where pairs of BT points are born. The contacts between Hopf and saddle-node curves are fold--Hopf, BT or nilpotent triple-zero points, and the barrier leaves only the first kind for $\Gamma<\frac14$.

The location of a BT point in the $(\omega,K)$ plane is a consequence of this structure rather than its starting point. The property of the balance that permits or forbids the degeneracy is the strength of carrier relaxation measured against the frequency of the relaxation oscillation; the injection and the detuning only decide where, within the permitted region, the degeneracy sits.

%======================================================================

\section*{Acknowledgements}
E. Chan-L\'opez (CVU 422090) and Carlos Antonio Mar\'in-Mendoza (CVU 714167) were supported by SECIHTI through the program ``Estancias Posdoctorales por M\'exico'' at the Divisi\'on Acad\'emica de Ciencias B\'asicas, Universidad Ju\'arez Aut\'onoma de Tabasco. A. Mart\'in-Ruiz acknowledges financial support from UNAM-PAPIIT (project IG100224), UNAM-PAPIME (project PE109226), SECIHTI (project CBF-2025-I-1862), and the Marcos Moshinsky Foundation. The authors thank Jaume Llibre for helpful comments and suggestions.

\appendix
\setcounter{table}{0}

\section{Algebraic derivation in the gain--loss frame}\label{app:A}

This appendix collects the algebra of the paper in one sequence. Each step uses only the preceding ones. We write $E=x+iy$, $I=|E|^{2}$, $\kappa=1+2Bn$, $s=\omega-\alpha n/2$ and $d=n^{2}/4+s^{2}$.

\subsection{Locked-state parametrization}\label{app:param}

For $E\neq0$, $\dot E=0$ gives $E=-K/(\frac12 n-is)$, hence
\begin{equation}\label{eq:A1}
x=-\frac{Kn}{2d},\qquad y=-\frac{Ks}{d},\qquad I=\frac{K^{2}}{d},
\end{equation}
and $\dot n=0$ gives $I=1-2\Gamma n/\kappa$ and $K^{2}=Id$. The map $(n,s)\mapsto(\omega,K)=(s+\alpha n/2,\sqrt{Id})$ is a bijection between $\{I(n)>0,\ \kappa(n)>0\}\setminus\{(0,0)\}$ and the set of locked states together with the injection parameters that sustain them. At fixed $(\omega,K)$ the locked inversions are the roots of $F(n)=I(n)[n^{2}/4+(\omega-\alpha n/2)^{2}]-K^{2}$, and $\kappa F$ is a cubic polynomial in $n$.

\subsection{Transport rate and characteristic invariants}\label{app:inv}

In the coordinates $(x,y,n)$ the Jacobian at a locked state is \eqref{eq:J}, with $\xi=\partial\dot n/\partial n=-2\Gamma-2B(I-1)$. On the balance law, $I-1=-2\Gamma n/\kappa$, so that
\begin{equation}\label{eq:A3}
\xi=-\frac{2\Gamma}{\kappa},\qquad I=1+\xi n,\qquad
\frac{dI}{dn}=\frac{\xi}{\kappa}.
\end{equation}
Every principal minor of $J$ that contains the $(3,3)$ entry is affine in it, with coefficient the complementary minor of $\widehat J$. Hence
\begin{equation}\label{eq:A4}
e_{1}=\dd_{1}+\xi,\qquad e_{2}=\xi\,\dd_{1}+\nn_{2},\qquad
e_{3}=\xi\,\dd_{2}+\nn_{3},
\end{equation}
with $\dd_{1}=\tr\widehat J=n$ and $\dd_{2}=\det\widehat J=d$. The two minors of $\widetilde J$ that contain the third row give $2\kappa[\frac12x(x-\alpha y)+\frac12y(\alpha x+y)]=\kappa I$, so $\nn_{2}=d+\kappa I$. Expanding $\det\widetilde J$ along the third row, the cross terms in $sxy$ cancel and
\begin{equation}\label{eq:A5}
\nn_{3}=\kappa\Bigl[\frac n2\,(x^{2}+y^{2})-\alpha s\,(x^{2}+y^{2})\Bigr]
=\kappa I\Bigl(\frac n2-\alpha s\Bigr).
\end{equation}
Using $\xi n=I-1$ and $d=K^{2}/I$ one obtains \eqref{eq:e23}. Since $\partial_{n}[n^{2}/4+(\omega-\alpha n/2)^{2}]=n/2-\alpha s$ and $dI/dn=\xi/\kappa$, the last expression is $e_{3}=\kappa\,\partial_{n}F$.

\subsection{Bogdanov--Takens conditions and the sum rule}\label{app:sum}

By \eqref{eq:e23} and $K^{2}=Id$,
\begin{equation}\label{eq:A7}
e_{2}=0\iff d=1-(1+\kappa)I\iff K^{2}=I\bigl[1-(1+\kappa)I\bigr].
\end{equation}
Because $n^{2}/4\le d<1$ and $d>0$, a BT point has $(1+\kappa)I<1$, $|n|<2$ and $K\le\frac12(1+\kappa)^{-1/2}$. The remaining condition $e_{3}=0$ is linear in $s$ once $d$ is fixed:
\begin{equation}\label{eq:A8}
\alpha\kappa I\,s=\xi d+\kappa I\,\frac n2 .
\end{equation}
Equations \eqref{eq:A7} and \eqref{eq:A8} are the whole BT system; the first contains no $\alpha$ and no $\omega$.

\subsection{Relaxation barrier}\label{app:barrier}

Eliminating $K$, $e_{2}=q(n)+s^{2}$ with $q$ as in \eqref{eq:e2q}. With $(1+\kappa)/\kappa-2=-2Bn/\kappa$,
\begin{equation}\label{eq:A10}
q(n)-\Bigl(\frac{n^{2}}{4}-4\Gamma n+1\Bigr)=2Bn+\frac{4\Gamma Bn^{2}}{\kappa}=\frac{2Bn\,(\kappa+2\Gamma n)}{\kappa}.
\end{equation}
For $n\ge0$ the right-hand side is nonnegative and $e_{2}\ge(n/2-4\Gamma)^{2}+1-16\Gamma^{2}\ge1-16\Gamma^{2}$. For $n<0$ the last term of $q$ is positive, so $e_{2}>n^{2}/4+\kappa\ge1-4B^{2}$, and in any case $e_{2}>0$. For $n>0$ the inequality $q(n)\le0$, necessary at a BT point, is equivalent to $\Gamma\ge\kappa(n^{2}/4+\kappa)/[2n(1+\kappa)]$, which gives $\Gamma_{\tan}(B)$ of Eq.~\eqref{eq:gtan}.

\subsection{Exact threshold for \texorpdfstring{$B=0$}{B=0}}\label{app:sharp}

For $B=0$, $\kappa=1$, $\xi=-2\Gamma$ and $I=1-2\Gamma n$. Then $e_{2}=s^{2}+n^{2}/4+2I-1=s^{2}-\sigma(n)$, and
\begin{equation}\label{eq:A12}
\begin{aligned}
e_{3}+2\Gamma e_{2}
&=-2\Gamma\Bigl(\frac{n^{2}}{4}+s^{2}\Bigr)+I\Bigl(\frac n2-\alpha s\Bigr)
  +2\Gamma\Bigl(s^{2}+\frac{n^{2}}{4}+2I-1\Bigr)\\
&=I\Bigl(\frac n2-\alpha s\Bigr)-2\Gamma(4\Gamma n-1)\\
&=I\bigl(g(n)-\alpha s\bigr),
\end{aligned}
\end{equation}
which is \eqref{eq:e3B0}; here we used $2I-1=1-4\Gamma n$ and the definition \eqref{eq:sigmag} of $g$. The BT points with $\alpha>0$ are the roots of $\Phi_{\alpha}=g^{2}-\alpha^{2}\sigma$ on $W_{\Gamma}=[n_{-},1/(2\Gamma))$, with $s=g/\alpha$; for $\alpha=0$ they are the zeros of $g$ on $W_{\Gamma}$, with $s=\pm\sigma^{1/2}$. Eliminating $s$ gives a quartic polynomial in $n$ after clearing the denominator $(1-2\Gamma n)^{2}$, so there are at most four candidates.

(i) \emph{The value $\Gc$.} At $n_{-}$ one has $4\Gamma n_{-}-1=n_{-}^{2}/4$, and
\begin{equation}\label{eq:A13}
g(n_{-})=\frac{n_{-}\,(1-3\Gamma n_{-})}{2\,(1-2\Gamma n_{-})}.
\end{equation}
Thus $g(n_{-})>0$ if and only if $n_{-}<1/(3\Gamma)$; since $1/(3\Gamma)<1/(2\Gamma)<n_{+}$, this holds if and only if $\sigma(1/(3\Gamma))=\frac13-1/(36\Gamma^{2})>0$, that is, $\Gamma>\Gc=1/(2\sqrt3)$, and $g(n_{-})=0$ exactly when $\Gamma=\Gc$. On the other hand,
\begin{equation}\label{eq:gquad}
g(n)=-\frac{2\Gamma n^{2}+(16\Gamma^{2}-1)\,n-4\Gamma}{2\,(1-2\Gamma n)} ,
\end{equation}
and the quadratic in the numerator has roots of product $-2$, hence exactly one positive root $n_{g}$; on $(0,1/(2\Gamma))$, $g>0$ for $n<n_{g}$ and $g<0$ for $n>n_{g}$. Three cases follow. If $\Gamma>\Gc$, then $g(n_{-})>0$, so $n_{g}\in W_{\Gamma}^{\circ}$, $Q(n_{g})=0$ and $\alpha_{0}=0$, attained. If $\Gamma=\Gc$, then $n_{g}=n_{-}=1/(3\Gc)$, $g<0$ on $W_{\Gamma}^{\circ}$ and $Q\to0$ as $n\to n_{-}^{+}$ (proof of Theorem~\ref{thm:sharp}), so $\alpha_{0}=0$, not attained in the interior; for $\alpha=0$ the solution is $n=n_{-}$, $s=0$, with $e_{1}=n_{-}-2\Gc=1/(6\Gc)\neq0$. If $\frac14<\Gamma<\Gc$, then $g(n_{-})<0$, so $n_{g}<n_{-}$ and $g<0$ on $W_{\Gamma}$; $Q$ is continuous and positive on $W_{\Gamma}^{\circ}$ and tends to $+\infty$ at both ends, so it attains a positive minimum and $\alpha_{0}>0$. Hence $\alpha_{0}(\Gamma)=0$ if and only if $\Gamma\ge\Gc$.

(ii) \emph{The limit $\Gamma\to\frac14^{+}$.} Finiteness of $\alpha_{0}$ holds because the quotient in \eqref{eq:alpha0} is finite at interior points. As $\Gamma\to\frac14^{+}$ the window shrinks to the point $n=2$; on it $1-2\Gamma n\to0$ and $4\Gamma n-1\to1$, so $|g|\to\infty$ uniformly, while $\sigma\le16\Gamma^{2}-1\to0$. Hence $\alpha_{0}\to\infty$.

(iii) \emph{Triple zero.} $e_{1}=0$ means $n=2\Gamma$; then $I=1-4\Gamma^{2}$, $d=8\Gamma^{2}-1$, $s^{2}=\sigma(2\Gamma)=7\Gamma^{2}-1$, and $\alpha s=g(2\Gamma)$ gives Eq.~\eqref{eq:TZ}; positivity of $I$ and of $s^{2}$ gives the range.

\subsection{Interval certificate for \texorpdfstring{$\mathrm{BT}_{1}$ and $\mathrm{BT}_{2}$}{BT1 and BT2}}\label{app:cert}

Table~\ref{tab:cert} lists the certificate used in Step~6 of Sec.~\ref{sec:example}, as produced by the verification script. The endpoints are exact decimal rationals; the signs of $\Phi_{2}$ are computed in exact rational arithmetic; the enclosures of $\Phi_{2}'$ are exact interval evaluations of the rational function $\Phi_{2}'$ on $[n_{\mathrm{lo}},n_{\mathrm{hi}}]$, rounded outward to ten decimals.

\begin{table}[ht]
\centering
\caption{Interval certificate for the roots of $\Phi_{2}$ ($B=0$, $\Gamma=\frac12$, $\alpha=2$). An exact Sturm count gives exactly two roots of the numerator of $\Phi_{2}$, with multiplicity, in $[4-2\sqrt3,1]$.}
\label{tab:cert}
\vspace{4pt}
\footnotesize
\setlength{\tabcolsep}{4pt}
\begin{tabular}{@{}lll@{}}
\toprule
 & $\mathrm{BT}_{1}$ & $\mathrm{BT}_{2}$\\
\midrule
$n_{\mathrm{lo}}$ & $0.53753419530643665778$ & $0.70893033878599019754$\\
$n_{\mathrm{hi}}$ & $0.53753419530643665779$ & $0.70893033878599019755$\\
$\sign\Phi_{2}(n_{\mathrm{lo}})$ & $+1$ & $-1$\\
$\sign\Phi_{2}(n_{\mathrm{hi}})$ & $-1$ & $+1$\\
$\Phi_{2}'$ enclosure & $[-7.8138834036,\,-7.8138834035]$ & $[17.8588837148,\,17.8588837149]$\\
zero excluded & yes & yes\\
$e_{1}=n-1$ enclosure & $[-0.46246580469356334222,$ & $[-0.29106966121400980246,$\\
 & $\phantom{[}{-}0.46246580469356334221]$ & $\phantom{[}{-}0.29106966121400980245]$\\
\bottomrule
\end{tabular}
\end{table}

\subsection{Normal-form coefficients}\label{app:normal}

At a BT point let $q_{0}$ and $p_{1}$ span the right and left kernels of $J$, and complete them by the Jordan chains $Jq_{1}=q_{0}$, $J^{\mathsf T}p_{0}=p_{1}$, normalized by $\langle p_{0},q_{0}\rangle=\langle p_{1},q_{1}\rangle=1$ and $\langle p_{0},q_{1}\rangle=\langle p_{1},q_{0}\rangle=0$. Because the first two rows and the first two columns of $J$ are independent ($\det\widehat J=d>0$), the kernel vectors are explicit: $q_{0}\propto J_{1\cdot}\times J_{2\cdot}$ and $p_{1}\propto J_{\cdot1}\times J_{\cdot2}$. Let $\mathcal B(u,v)$ be the bilinear second derivative of the vector field; its components are
\begin{equation}\label{eq:A16}
\begin{aligned}
\mathcal B_{1}&=\tfrac12(u_{3}v_{1}+u_{1}v_{3})-\tfrac{\alpha}{2}(u_{3}v_{2}+u_{2}v_{3}),\\
\mathcal B_{2}&=\tfrac{\alpha}{2}(u_{3}v_{1}+u_{1}v_{3})+\tfrac12(u_{3}v_{2}+u_{2}v_{3}),\\
\mathcal B_{3}&=-2\kappa(u_{1}v_{1}+u_{2}v_{2})-4B\,[x(u_{1}v_{3}+u_{3}v_{1})+y(u_{2}v_{3}+u_{3}v_{2})].
\end{aligned}
\end{equation}
The coefficients of the normal form $\dot w_{0}=w_{1}$, $\dot w_{1}=aw_{0}^{2}+bw_{0}w_{1}$ are \cite[Sec.~8.7]{Kuznetsov2004}
\begin{equation}\label{eq:A17}
a=\tfrac12\langle p_{1},\mathcal B(q_{0},q_{0})\rangle,\qquad
b=\langle p_{0},\mathcal B(q_{0},q_{0})\rangle+\langle p_{1},\mathcal B(q_{0},q_{1})\rangle .
\end{equation}
Replacing $q_{0}$ by $cq_{0}$ multiplies $(a,b)$ by $c$, and the remaining freedom $q_{1}\mapsto q_{1}+tq_{0}$, $p_{0}\mapsto p_{0}-tp_{1}$ leaves $b$ unchanged, so the invariant content is $ab\neq0$ and $\sign(ab)$. In the normal form, $\sign(ab)=-1$ corresponds to a supercritical Hopf curve emanating from the BT point and $\sign(ab)=+1$ to a subcritical one, for the dynamics on the center manifold \cite{Kuznetsov2004}.

\bibliographystyle{elsarticle-num}
\bibliography{CarrierRelaxation_BogdanovTakens_CSF}

\end{document}